\documentclass[12pt,preprint]{aastex}

\usepackage{natbib}
\usepackage{rotate}
\usepackage{lscape}
\def\gtorder{\mathrel{\raise.3ex\hbox{$>$}\mkern-14mu
             \lower0.6ex\hbox{$\sim$}}}
\def\ltorder{\mathrel{\raise.3ex\hbox{$<$}\mkern-14mu
             \lower0.6ex\hbox{$\sim$}}}

\def\bfy{{\bf y}}
\def\bfn{{\bf n}}
\def\bfs{{\bf s}}
\def\bfq{{\bf q}}
\def\bfp{{\bf p}}

\def\bfhs{{\bf\hat{s}}}
\def\bfhq{{\bf\hat{q}}}
\def\bfds{{\bf\Delta s}}
\def\bfdq{{\bf\Delta q}}

\shorttitle{Quantifying Quasar Variability}
\shortauthors{Koz{\l}owski et al.}

\begin{document}

\title{Quantifying Quasar Variability As Part of a General Approach To Classifying Continuously Varying Sources}

\author{Szymon~Koz{\l}owski\altaffilmark{1}, 
Christopher~S.~Kochanek\altaffilmark{1,2}\\
and\\
A.~Udalski\altaffilmark{3},
{\L}.~Wyrzykowski\altaffilmark{3,4},
I.~Soszy{\'n}ski\altaffilmark{3},
M.~K.~Szyma{\'n}ski\altaffilmark{3}, 
M.~Kubiak\altaffilmark{3},\\
G.~Pietrzy{\'n}ski\altaffilmark{3,5},
O.~Szewczyk\altaffilmark{3,5},
K.~Ulaczyk\altaffilmark{3}, 
R.~Poleski\altaffilmark{3}\\ (The OGLE collaboration)
}

\altaffiltext{1}{Department of Astronomy, The Ohio State University, 140 West 18th Avenue, Columbus, OH
43210, USA; simkoz@astronomy.ohio-state.edu; ckochanek@astronomy.ohio-state.edu}
\altaffiltext{2}{The Center for Cosmology and Astroparticle Physics, The Ohio State University, 
191 West Woodruff Avenue, Columbus, OH 43210, USA}
\altaffiltext{3}{Warsaw University Observatory, Al. Ujazdowskie 4, 00-478 Warszawa, Poland}
\altaffiltext{4}{Institute of Astronomy, University of Cambridge, Madingley Road, Cambridge CB3 0HA, UK}
\altaffiltext{5}{Universidad de Concepci{\'o}n, Departamento de Fisica, Casilla 160-C, Concepci{\'o}n, Chile}

\begin{abstract}
Robust fast methods to classify variable light curves in large sky surveys 
are becoming increasingly important. While it is relatively straightforward to identify common periodic stars and 
particular transient events (supernovae, novae, microlensing events), there is no equivalent for
non-periodic continuously varying sources (quasars, aperiodic stellar variability).
In this paper we present a fast method for modeling and classifying such sources.
We demonstrate the method using $\sim 86,000$ variable sources from the OGLE-II survey 
of the LMC and $\sim 2,700$ mid-IR selected quasar candidates from the OGLE-III survey of the LMC and SMC. 
We discuss the location of common variability classes in the parameter
space of the model. In particular, we show that quasars
occupy a distinct region of variability space, providing a simple quantitative approach to the variability
selection of quasars. 

\end{abstract}


\section{Introduction}

The focus of astronomy on the transient universe is steadily increasing because
of the wealth of astrophysical information provided by time variability.  Examples
of past and current variability surveys are the Galactic microlensing surveys 
(e.g., OGLE, \citealt{1997AcA....47..319U}; MOA, \citealt{2003ApJ...591..204S}; 
MACHO, \citealt{2000ApJ...542..281A}; EROS, \citealt{1995A&A...303..137B}; 
SuperMACHO, \citealt{2005IAUS..225..357B}), searches for both
local (LOSS, \citealt{2001ASPC..246..121F}) and cosmologically distant 
(e.g., SDSS, \citealt{2008AJ....135..348S}; CFHTLS, \citealt{2007A&A...461..813C}; 
Essence, \citealt{2007ApJ...666..674M}) supernovae, all sky variability surveys (e.g., ASAS,
\citealt{1997AcA....47..467P}; QUEST, \citealt{2004AJ....127.1158V}; 
NSVS, \citealt{2004AJ....127.2436W}; Catalina, \citealt{2009ApJ...696..870D}) and prompt $\gamma$-ray
burst monitors (e.g., BATSE, \citealt{1992Natur.355..143M}; RAPTOR, \citealt{2002SPIE.4845..126V}; ROTSE, \citealt{2005ApJ...631L.121R}; 
SWIFT, \citealt{2005SSRv..120..165B}).  A complete, or even partial, listing of more directed
monitoring projects would be impossible, and the scale of variability surveys
continues to grow rapidly with projects such as OGLE-IV (A. Udalski et al. 2010, in preparation), 
Pan-STARRS (\citealt{2004AN....325..636H}) and LSST (\citealt{2008arXiv0805.2366I}).

A common problem for variability surveys is the classification of light curves given
the enormous range of physical processes leading to variability (e.g., \citealt{1996lcvs.book.....S}).  
Classification is well developed for simple periodic sources (e.g., Cepheids, RR Lyrae, 
eclipsing binaries, etc.) and impulsive (e.g., microlensing) or explosive sources 
(e.g., supernovae, novae), but less well developed for sources with non-periodic, 
continuous or very long time-scale variability such as quasars.  Most efforts
at classification have focused on the broad spectrum of periodic sources, 
using the Fourier amplitudes of the light curves to recognize different
variable classes using a broad range of methods (e.g.,
\citealt{2002AcA....52..241E,2002AcA....52..397P, 2005MNRAS.358...30E, 2008AcA....58..163S}).
More sophisticated methods for classification of variable objects include neural networks (e.g., \citealt{2004MNRAS.352..233B})
and self-organizing maps (e.g., \citealt{2008AIPC.1082..201W}).

Attempts at classifying non-periodic, continuously variable sources have
largely focused on identifying quasars. \cite{2007AJ....134.2236S} found that
essentially all quasars vary at some level on long time scales, with 90\% of
Sloan Digital Sky Survey (SDSS) quasars having rms variability $\geq0.03$ mag on a 6 year time scale. The average variability of 
quasars can be described by a structure function with greater variability
amplitudes on longer time baselines (e.g., \citealt{2004ApJ...601..692V,2005AJ....129..615D}), and this has
also been observed for smaller numbers of individually monitored 
quasars (e.g., \citealt{1985ApJ...296..423C,1996ApJ...465..87N,2002A&A...388..771C,2005ApJ...622..129S}). 
Thus, variability searches for quasars (e.g., \citealt{1983MNRAS.202..571H};
\citealt{1995A&A...296..665V}, or 
\citealt{2003AJ....125.1330D,
2003AJ....125....1G,2004ApJ...606..741R,2005A&A...442..495D,2005MNRAS.356..331S} more recently)
generally look for non-periodic sources with long time scale variability.
The most formal approach to date is that used by Eyer (2002) and Sumi
et al. (2005), where the selection criterion was based on the slope of
the structure function of the light curves.  These searches have generally
been more successful in relatively empty extragalactic regions than in
those with high stellar densities.  

A significant problem in these attempts is that there was no simple method
to reduce a non-periodic, continuously variable light curve to a small set of
numbers, such as the periods and amplitudes of a Fourier series, which can be used
for classification.  Recently, \cite{2009ApJ...698..895K} found that quasar light
curves could be well modeled as a stochastic process with an exponential 
covariance matrix characterized by an amplitude and a time scale based on
a small sample ($\sim 100$) of quasars with multi-year light curves.  Physically,
the model is a damped random walk, and it has a broken power law structure function
consistent with studies of quasar structure functions.
Unfortunately, there are few large samples of quasars with extensive monitoring data. To our knowledge,
there are the quasars in the SDSS equatorial strip, with roughly 60 epochs
over $\sim$six years (\citealt{2007AJ....134.2236S,2008MNRAS.386..887B}), the QUEST survey (\citealt{2004ApJ...617..184R}), 
whose light curves are not publicly available, and any quasars
lying in the microlensing survey regions.  Recently, \cite{2009ApJ...701..508K}
used mid-IR color selection (e.g., \citealt{2005ApJ...631..163S}, also see \citealt{2004ApJS..154..166L}) 
to identify a large sample of quasar candidates in the Magellanic Clouds.  While they have yet to be 
spectroscopically confirmed, the nature of the selection method means that 
purity of the main sample of candidates should be high.  By combining
these candidates with the OGLE-II and OGLE-III (\citealt{1997AcA....47..319U,2008AcA....58...69U}) light curves, 
we can verify that quasar light curves are well described by a stochastic process
based on a much larger sample of objects than used by \cite{2009ApJ...698..895K}
and then explore using this approach to light curve modeling for classifying
light curves in general, including periodic sources.  Obviously, a damped random 
walk is not an optimal model for periodic sources, but there is no technical 
difficulty in modeling them using the stochastic process.

This paper is organized as follows. In Section~\ref{sec:data} we describe the light curve data
and the existing catalogs of variable sources we can use for classifying
light curves.  In Section~\ref{sec:method} we describe the mathematical model.
We do not use the particular forecasting methodology of \cite{2009ApJ...698..895K},
but a more statistically optimal approach based on \cite{1992ApJ...385..404P} and 
\cite{1992ApJ...398..169R,1994comp.gas..5004R} that we detail in the appendix.  In
Section~\ref{sec:quasars} we examine how well the model works for the known quasars from \cite{2009ApJ...698..895K}
and the \cite{2009ApJ...701..508K} quasar candidates, the distribution of quasars
in the parameter space of the model and the completeness of variability selection.  
In Section~\ref{sec:genvar} we examine the problem of contamination by determining the
distribution of the general population of variable LMC sources in the model parameters.  
We also examine how the stochastic process models treat true periodic variables.  
In Section~\ref{sec:selection_quasars} we combine these results and discuss the problem 
of variability selecting quasars in dense stellar fields like the LMC.
We summarize our results and discuss future uses in Section~\ref{sec:summary}.


\section{Data}
\label{sec:data}

The data used in this paper were collected with the 1.3 m Warsaw telescope at the Las 
Campanas Observatory, Chile during the second (1996--2000) and third (2001--2008) phases of 
the OGLE project (\citealt{1997AcA....47..319U,2008AcA....58...69U}). The majority of the OGLE data were 
collected in the $I$-band ($\sim$ 370 points per light curve in OGLE-III) with a smaller number of points 
in the $V$-band ($\sim$ 40 points per light curve). The OGLE-III light curves for many common variable stars are 
available on the internet\footnote{\tt http://ogledb.astrouw.edu.pl/$\sim$ogle/CVS/}.
The OGLE-II database of light curves ($I$-band only) can be accessed through several web interfaces\footnote{{\tt http://ogledb.astrouw.edu.pl/$\sim$ogle/photdb/} and \tt http://ogle.astrouw.edu.pl/ogle2/dia/} 
(\citealt{2005AcA....55...43S,1997AcA....47..319U}).  
We make use of published OGLE samples of classical 
Cepheids (Cep; \citealt{2008AcA....58..163S}), double mode Cepheids 
(dCep; \citealt{2008AcA....58..153S}), RR Lyrae stars (RRLyr; \citealt{2009AcA....59....1S}), 
OGLE Small Amplitude Variable Red Giants (OSARGs; \citealt{2004AcA....54..129S}), 
eclipsing binaries\footnote{\tt http://ogle.astrouw.edu.pl/ogle2/lmc\_ecl/} (ECL; \citealt{2003AcA....53....1W}), 
ellipsoidal variable red giants\footnote{\tt http://ogle.astrouw.edu.pl/cont/4\_main/var/ogleii/ell/ell.html} 
(ELL; \citealt{2004AcA....54..347S}), long secondary period variables (LSPs; \citealt{2007ApJ...660.1486S}), and
long period variables (Miras, LPVs, and other semiregular variables, \citealt{2005AcA....55..331S}) in the LMC.
We separately extracted the OGLE-II light curves of Be 
stars from \cite{2002AJ....124.2039K}, $\sim$ 300 OGLE-II and $\sim$ 2700 OGLE-III 
light curves of the mid-IR selected quasar candidates from \cite{2009ApJ...701..508K}. 
We also reanalyzed the 109 quasars from \cite{2009ApJ...698..895K}.

In addition to these specific samples we downloaded $\sim$ 150,000 ``variable objects'' from six 
OGLE-II LMC ``inner'' fields (SC1 -- SC6) covering about 1.3 deg$^2$ and centered on the highest stellar 
density regions, and $\sim$ 38,000 ``variable objects'' from the three lowest stellar density ``outer'' fields (SC15, SC19, and SC20).
The core of this paper focuses on the analysis of the highest density inner fields as a worst case scenario
for identifying quasars, with the lower stellar density outer LMC fields as a simpler comparison sample.
An object was defined to be variable if the standard deviation of its light curve from a constant flux was 
at least twice the median photometric error (hereafter called the variability 
criterion). A large fraction of these ``variables'' are flat light curves with small numbers of 
outliers, and these are removed by the following cleaning procedure. First, 
we correct the photometric errors following the procedure and values from \cite{2009arXiv0905.2044W}. 
Then we removed up to five significant ($>5\sigma$) outliers (relative to their adjacent epochs)
from each light curve, where a typical OGLE-II light curve has 360 epochs. 
After removing the sources no longer satisfying the variability 
criterion, we are left with 86,301 light curves in the inner LMC fields and 25,956 in the outer ones. 
The inner fields include 584 RR Lyrae (499
``ab'' and 85 ``c''), 133 classical Cepheids, 7 double mode Cepheids, 6,745 OSARGs, 374 ellipsoidal 
variable red giants, 46 eclipsing binaries, 699 LSPs, 1,442 LPVs and 15 Be stars. We also use the
58 OGLE-II and 984 OGLE-III quasar candidates meeting the variability criterion. 

On further inspection many of these ``variable'' stars are created by problems in the wings
of the point spread functions (PSFs) of bright (variable) stars. This has been noted in earlier studies (e.g., \citealt{2002AcA....52..241E,2005MNRAS.356..331S}).
In particular, faint stars near bright variable stars tend to vary in phase 
with the bright source. We apply a three step procedure to remove these ``ghost'' variable objects. 
First, we mask regions near bright stars. Stars brighter than $I\approx12.5$ mag are saturated in the OGLE data. 
These stars have PSF wings extending to 60 arcsec but they are not formally ``detected'' in the OGLE catalogs. 
In order to mask these stars, we used stellar objects from the {\sc 2mass} 
catalog (\citealt{2003tmc..book.....C}) brighter than $J=14$ mag. The masks are defined by 
rectangular regions with $\Delta \rm R.A.=2\delta$, $\Delta \rm decl.=\delta$ around the Two Micron All Sky Survey
({\sc 2mass}) position of each star, where $\delta = 120(1 - J/14)$ arcsec.
If a variable object is inside one of these regions, it is  
removed from the list of variables. We also masked visually identified regions which showed clear
photometric problems (a dense clump of variable objects, variables detected on a ``bloomed'' CCD line, etc.).
Our second criterion is that no variable could have any other variable object within 6 arcsec.
Finally, the third criterion is that no variable can have a $>2$~mag brighter variable within 15 arcsec. 
These steps could be carried out with greater precision, but cleaning the OGLE catalog of spurious variables is not
our primary focus even if it is ultimately a limitation and is one reason for using our rather conservative 
variability criterion. One additional problem with the OGLE-II
photometry is a result of the realuminization of the telescope's mirror. A small number of light curves show a step-like 
change in magnitude between seasons 2 and 3. These light curves are identified during the visual inspection we carried out for the 
final selection of quasar candidates. A summary of these photometric problems is also given in 
\cite{2002AcA....52..241E} and \cite{2005MNRAS.356..331S}.


\section{Methodology}
\label{sec:method}

We are interested in modeling random processes leading to variability.
Many such processes can be described by the covariance matrix $S$ of the signal,
where we will be considering the particular case of an exponential
covariance 
\begin{equation}
        S_{ij} = \sigma^2 \exp(-|t_i-t_j|/\tau)
 \label{eqn:cfunc}
\end{equation}
between epochs at $t_i$ and $t_j$
used by \cite{2009ApJ...698..895K} to model quasar light curves based on a
standard method from the forecasting literature (see, e.g., \citealt{BD2003}). 
As explained in detail by \cite{2009ApJ...698..895K},
the physics of this process is a random walk driving term of amplitude
$\sigma$ balanced by a damping time scale $\tau$ for a return to the mean.
It is useful to define $\sigma^2 = \tau \hat{\sigma}^2/2$ 
and use $\tau$ and $\hat{\sigma}$ as our model parameters -- on short
time scales $|\Delta t| << \tau$ the dispersion between two points is
$\hat{\sigma}|\Delta t|^{1/2}$ while on long time scales it asymptotes
to $\sigma$.   Computationally, there is less covariance between
$\tau$ and $\hat{\sigma}$ than $\tau$ and $\sigma$. The power spectrum 
of the process is
 \begin{equation}
        P(f) = \frac{2\hat{\sigma}^2\tau^2}{1+(2\pi \tau f)^2}
 \label{eqn:Pf}
\end{equation}
corresponding to white noise on long time scales $(f\rightarrow0)$
and then a $\propto f^{-2}$ fall on short time scales $(f>1/\tau)$.
\cite{2009ApJ...698..895K} suggest that $\tau$ may be related to the thermal
time scale of accretion disks. We estimate these parameters
using the approach of \cite{1992ApJ...385..404P}, its generalization 
in \cite{1992ApJ...398..169R} and the fast computational implementation 
in \cite{1994comp.gas..5004R}.  We will collectively refer to these as 
the PRH method.  In the appendix we 
summarize this approach, show how it can be used to derive
the forecasting model of \cite{2009ApJ...698..895K}, and 
demonstrate that it is more statistically powerful than the
forecasting approach while still requiring only $O(N)$ operations
for a light curve with $N$ points.

Operationally, we fit a light curve by maximizing the likelihood (Equation (\ref{eqn:likefit}))
that the light curve can be fitted by the process model to determine the
parameters $\tau$ and $\hat{\sigma}$ (as well as their uncertainties, if
desired).  We include the light curve mean as a simultaneously optimized
linear parameter (Equations (\ref{eq:A4}) and (\ref{eq:A7})).  Given the estimates for
$\tau$ and $\hat{\sigma}$, we have optimal estimates for the mean light curve
at both the observed points (Equation (\ref{eqn:predict})) and at any other time (Equation (\ref{eq:A11})),
as well as the variances about these means (Equations (\ref{eq:A9}) and (\ref{eq:A12}), respectively).
Where we show these reconstructions and the ``error snakes'' defined by
the variances, there are two important points to consider when comparing
them to the data points.  First, these variances are the variances in
the mean light curve and not the variance of the data relative to the
mean light curve.  The latter quantity, given in Equation (\ref{eqn:var1}), is defined
only where there is data and so is ill-suited to showing a continuous
light curve.  Data points will be scattered relative to the mean light
curve by the combination of the variance in the mean light curve and the uncertainties in
the individual data points.  Second, the reconstructed light curve is not an example of
an individual random walk defined by the parameters ($\tau$ and $\hat{\sigma}$),
but rather the average of all such random walks that are consistent with
the observed light curve given its uncertainties.  The variance in the
reconstructed light curve is then the variance of these individual random
walks about the mean.  If we generated individual random walks constrained
by the data (see PRH), they would track the mean light curve and (statistically)
stay inside the ``error snake'' defined by the variances, but they would
show more structure on short time scales and excursions outside the ``error snake''
consistent with the estimated variances.

When we fit the model, we obtain a maximum likelihood estimate of $\tau$ and $\hat{\sigma}$
with log likelihood $\ln L_{best}$. There are two limiting cases for this model.
First, as the time scale $\tau\rightarrow0$, the covariance matrix 
becomes a diagonal matrix $S_{ij}\rightarrow\sigma^2\delta_{ij}$ equivalent
to simply broadening the photometric errors. We refer to this limit as the 
``noise'' limit with log likelihood $\ln L_{noise}$.  We use a cut of $\ln L_{best} > \ln L_{noise}+2$ to 
select objects that are modeled by the exponential process. Given the variability
criterion from Section~2, this cut mainly eliminates variable objects with $\tau$ smaller 
than the mean epoch separation.  The other limit
$\tau\rightarrow\infty$, with likelihood $\ln L_\infty$, is the limit where we cannot determine
$\tau$ given the overall time span of the survey. This has not proved to be major 
issue with the OGLE data.  We cannot distinguish $\tau$ from $\tau \rightarrow \infty$ 
for only 2\% of the $\sim 86,000$ OGLE-II variables and 4.5\% of the OGLE-III quasar candidates. 
For these objects the time scale $\tau$ was set to $\log_{10}(\tau)\simeq5$.

We also check for periodicity in each light curve using the Lomb-Scargle periodogram 
method as implemented by \cite{1992nrfa.book.....P}. For each light curve we note the three 
most likely periods $P_i$ $(i=1,2,3)$ and their probabilities. We used narrow notches
at 1, 1/2, 1/3 and 1/4 of a day to minimize diurnal aliasing problems,
while still being able to examine RR Lyrae periods.
These aliasing solutions could be improved, but this 
is not the focus of this paper. We classify a source as periodic if the estimated period
$P$ is shorter than 200 days and the likelihood for the peak in the periodogram being
observed at random is $\log_{10}(p_{\rm periodic})<-3$.   We include a limit that periods
must be shorter than 200 days because many quasars with large $\tau$ 
have a ``period'' satisfying $\log_{10}(p_{\rm periodic})<-3$ with $P>200$~day. These solutions corresponds
to fitting a sine wave to the data, and as the number of time scales $\tau$ covered
by the light curves shrinks, there is an increasing likelihood that a sine
wave of some period will enormously improve the goodness of fit over no variability,
leading to a false positive in the periodogram.
For periodic variables we should see a strong correlation between $P$ and $\tau$.


\section{Quasars}
\label{sec:quasars}

For our quasar samples we use the 109 known quasars from \cite{2009ApJ...698..895K}
and $\sim 2,700$  mid-IR selected quasar candidates detected in the OGLE-III LMC and SMC fields from \cite{2009ApJ...701..508K}. 
The \cite{2009ApJ...698..895K} sample is a heterogeneous mixture of MACHO variability selected
quasars (\citealt{2003AJ....125....1G}) and quasars monitored at least in part for reverberation mapping
(\citealt{1999MNRAS.306..637G,2004ApJ...613..682P}). They have the advantage of being confirmed
quasars, but they are a non-random sample of quasars, and many of the reverberation mapping 
quasars have poorly sampled light curves compared to the MACHO or OGLE light curves. 
The mid-IR selected candidates have the disadvantage that they are not confirmed quasars.
\cite{2009ApJ...701..508K} classified the candidates based on their mid-IR colors
(A/B for being inconsistent/consistent with a black body),
location in the mid-IR color-magnitude diagram (CMD; YSO/QSO for whether the object is/is not in 
the region contaminated with young stellar objects, YSOs)
and optical to mid-IR color (a/b for whether the optical to mid-IR colors are/are not consistent with 
spectroscopically selected quasars).  Here we only use the ``a'' sources. 
Based on the extragalactic sources from the AGN and 
Galaxy Evolution Survey (AGES, C. S. Kochanek et al. 2010, in preparation), the QSO-Aa candidates, which
are the majority, should be almost entirely comprised of quasars, while the other classes are likely dominated 
by contaminating sources.

Ideally, we should work with a quasar light curve that is free from contamination by the
host galaxy. This can be a problem because we fit the light curves in magnitudes rather than flux (as did \citealt{2009ApJ...698..895K}).
In magnitude space, the effects of contamination on the process parameters are non-linear. Extra contaminating flux
reduces amplitudes and increases time scales, while oversubtracting any contamination has
the reverse effects. For the bright, distant quasars from \cite{2003AJ....125....1G}, 
\cite{2009ApJ...701..508K} and, to a large extent, \cite{1999MNRAS.306..637G}, this
will be a limited problem.  Not only are these very luminous quasars, but the 
observations are largely in the rest frame ultraviolet, where host galaxies have little flux. 
 The \cite{2009ApJ...701..508K} sample will 
also have few sources with significant host contamination because such sources are lost as 
part of the mid-IR selection (see \citealt{2008ApJ...679.1040G,2009arXiv0909.3849A}).
It is emphatically 
not true of the local reverberation mapping targets.  If we take the continuum
light curves for MRK~279 \citep{2001A&A...369...57S} , MRK~509 \citep{1996ApJ...471..737C}, 
NGC~3783 \citep{1994ApJ...425..609S}, NGC~4051 \citep{2000ApJ...542..161P},
and NGC~5548 \citep[][and references therein]{2002ApJ...581..197P},
and then either subtract the remaining host flux in the spectroscopic apertures or add in the host flux from outside
the spectroscopic apertures based on \cite{2009ApJ...697..160B}, we see significant changes
in the model parameters.   On average, adding the rest of the host flux (subtracting the remaining
host flux in the aperture) reduces (increases) the variability amplitude by an average of
$-0.73$~dex ($+0.38$~dex) and increases (decreases) the 
characteristic time scale by an average of $+0.77$~dex ($-0.42$~dex) for five sources. 
The effects are larger when including the remainder of the host because the spectroscopic
apertures used to obtain these light curves are already excluding most of the host galaxy flux -- the host flux outside
the aperture is 1--10 times larger than the host flux in the aperture.  Thus, as
with reverberation mapping, correlations of the process parameters with other
quasar properties will be sensitive to the treatment of the host galaxy.  We will not
consider this problem further since we are focusing on the observed properties of
quasar variability rather than their detailed interpretation.

\begin{figure}[t]
\centering
\includegraphics[width=16cm]{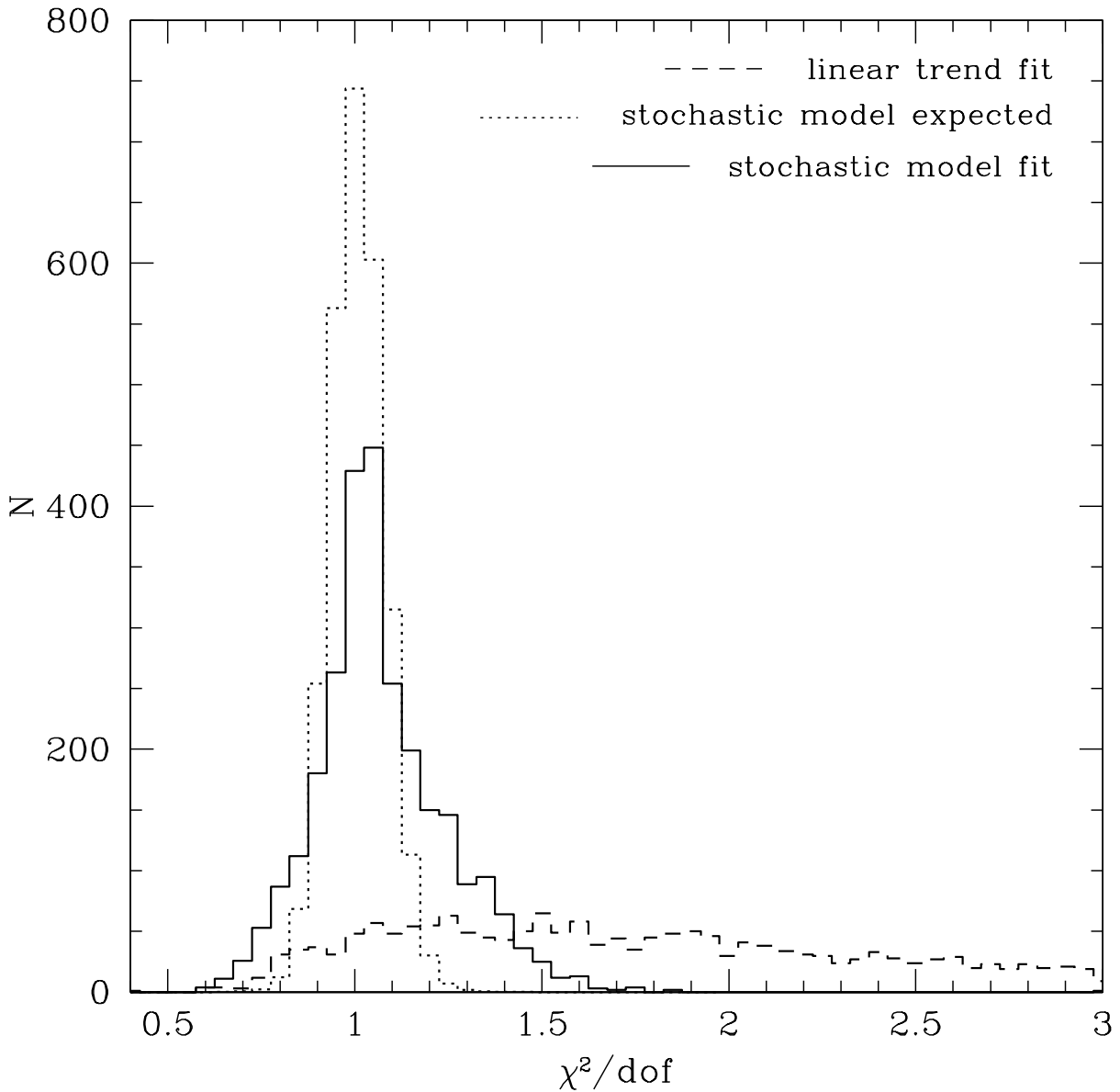}
\caption{Distributions of the goodness of fit for the stochastic model (linear trend)
are shown by the solid (dashed) line for the $\sim 2,700$ OGLE-III light curves of the mid-IR selected quasars from \cite{2009ApJ...701..508K}. The improvement over fitting the linear trend from using the stochastic model is clearly visible. The dotted line is the expected distribution of $\chi^2/dof$  based on the number of degrees of freedom for each light curve.  The differences between the stochastic model and the expected distribution are some combination of errors in the estimated photometric errors, small numbers of outliers that have not been eliminated from the light curves and any poorly modeled physics.
}
\label{fig:chi2dof}
\end{figure}

Figure~\ref{fig:four_qso} shows four examples of QSO light curves modeled by the stochastic process.
These fits are typical, and like \cite{2009ApJ...698..895K}, we find that most quasars are well modeled 
by the process, as illustrated by the $\chi^2/dof$ distribution shown in Figure~\ref{fig:chi2dof}. 
Some care is required to interpret this distribution. All objects we consider are likely
to be poorly described by no variability, so the process model will always produce a reduced $\chi^2$.
Also keep in mind that one limit of the model is simply to broaden the error bars (the ``noise'' limit),
which can always allow $\chi^2/dof \rightarrow 1$ up to the effects of the priors on 
$\hat{\sigma}$ and $\tau$. Nonetheless, the final $\chi^2/dof$ distribution is far narrower
than the distribution from fitting the linear model and few objects are doing so
in the ``noise limit'' corresponding to simply broadening the uncertainties.
The distribution is somewhat broader than it should
be for correctly estimated Gaussian uncertainties.  For a typical OGLE-III light 
curve with $N=360$ points, we expect $\chi^2/N\simeq1\pm\sqrt{2/N}=1\pm0.07$.  That
the distribution is broader, and relatively symmetric, suggests that some of the
difference lies in the accuracy of the photometric errors, since a 3\% misestimation
of the uncertainties is enough to cause such a shift.  There is some skewness
to larger $\chi^2/dof$, but this could simply be due our pruning of outliers
being too conservative -- it only takes two  $4\sigma$ outliers to
produce an $0.07$ shift in $\chi^2/N$.  
Nonetheless, we cannot rule out the possibility that
the process model is a poor or incomplete model for the variability physics in
some objects.  

\begin{figure}[t]
\centering
\includegraphics[width=16cm]{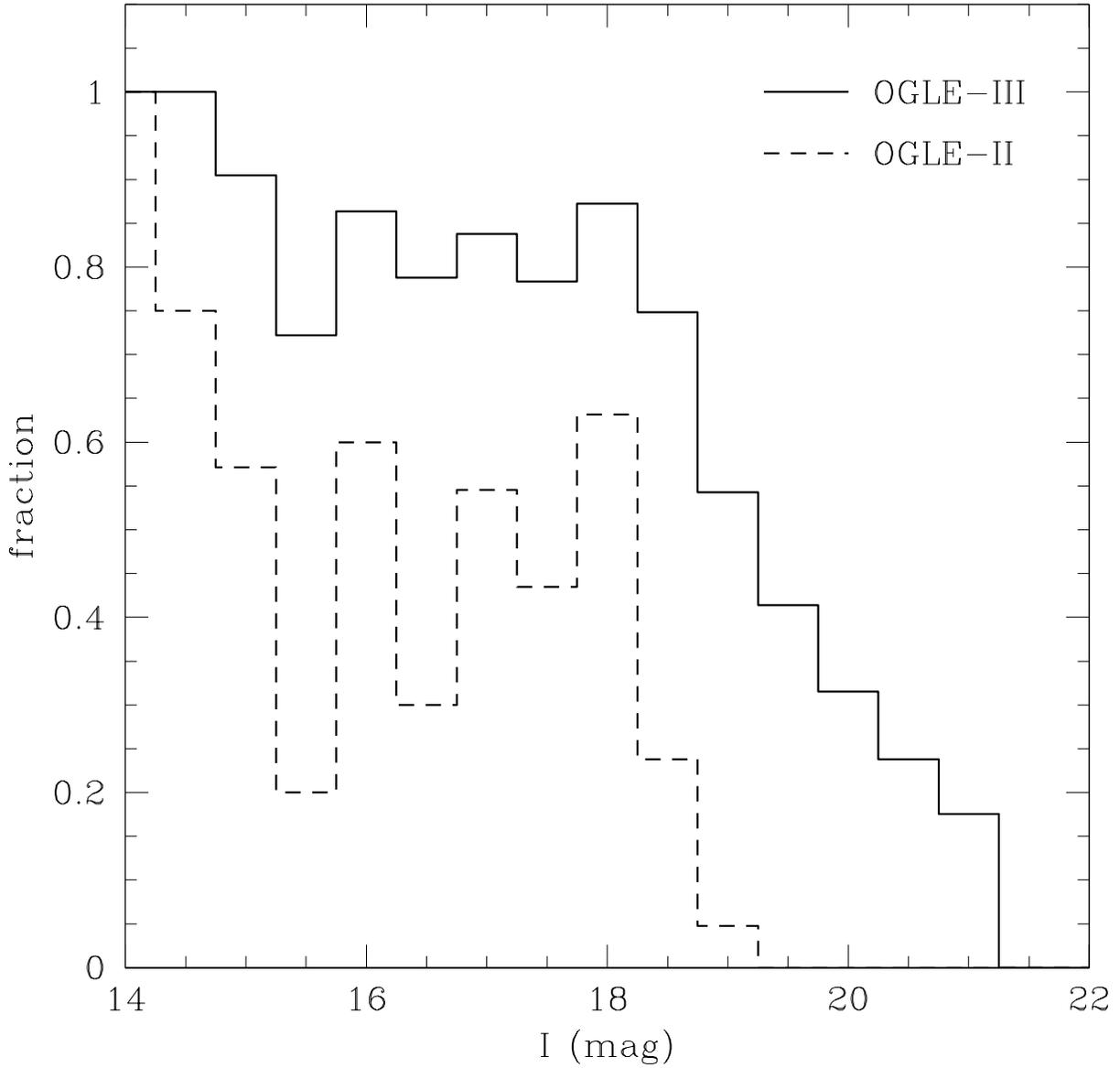} 
\caption{Fraction of variable QSO and YSO candidates as a function of magnitude for OGLE-III (solid line) and OGLE-II (dashed line). A source is counted as variable if it passes the variability criterion of Section~\ref{sec:data} and has $\ln L_{best} > \ln L_{noise}+2$.}
\label{fig:frac_var}
\end{figure}

The detection of variability is strongly magnitude-dependent because 
fainter sources require larger amplitudes for variability to be detected.
Figure~\ref{fig:frac_var} shows the fraction of 
mid-IR quasar candidates that pass the two variability criteria based on either 
OGLE-II or OGLE-III data.  One advantage of the process model is that we can 
explicitly calculate the completeness of a survey as a function of the parameters
using Monte Carlo simulations.   At each point on a grid of the process parameters
($\hat{\sigma}$ and $\tau$), we randomly generate $N$ 
light curves using the temporal sampling of a randomly selected real light curve.
Generating simulated light curves of zero mean is trivial, since the next point
\begin{equation}
s_{i+1}=s_i e^{-\Delta t/\tau} + G\left[\sigma^2\left(1-e^{-2\Delta t/\tau}\right)\right],
\label{eqn:nextpoint}
\end{equation}
where $\Delta t$ is the time interval and $G(x^2)$ is a Gaussian deviate of
dispersion $x$. The observed light curve is then $y_i=s_i+G(n_i^2)$ where $n_i$
is the observational noise. The chain is initiated by $s_1=G(\sigma^2)$.
We use the mean measurement error at a given magnitude, but one could
use the error estimates from a light curve matched in magnitude to the trial.
Next, we apply our variability criterion, that
the rms of the light curve must be twice the measurement errors,
and that $\ln L_{best} > \ln L_{noise}+2$, to each trial
light curve. The completeness is simply the fraction of the trials which
pass both selection criteria.  For speed, we estimate the second criterion
by using the likelihood at the input parameters for $L_{best}$ and the
likelihood for the model with the same asymptotic variance but $\tau\rightarrow 0$
for $L_{noise}$.

\begin{figure}[t]
\centering
\includegraphics[width=16cm]{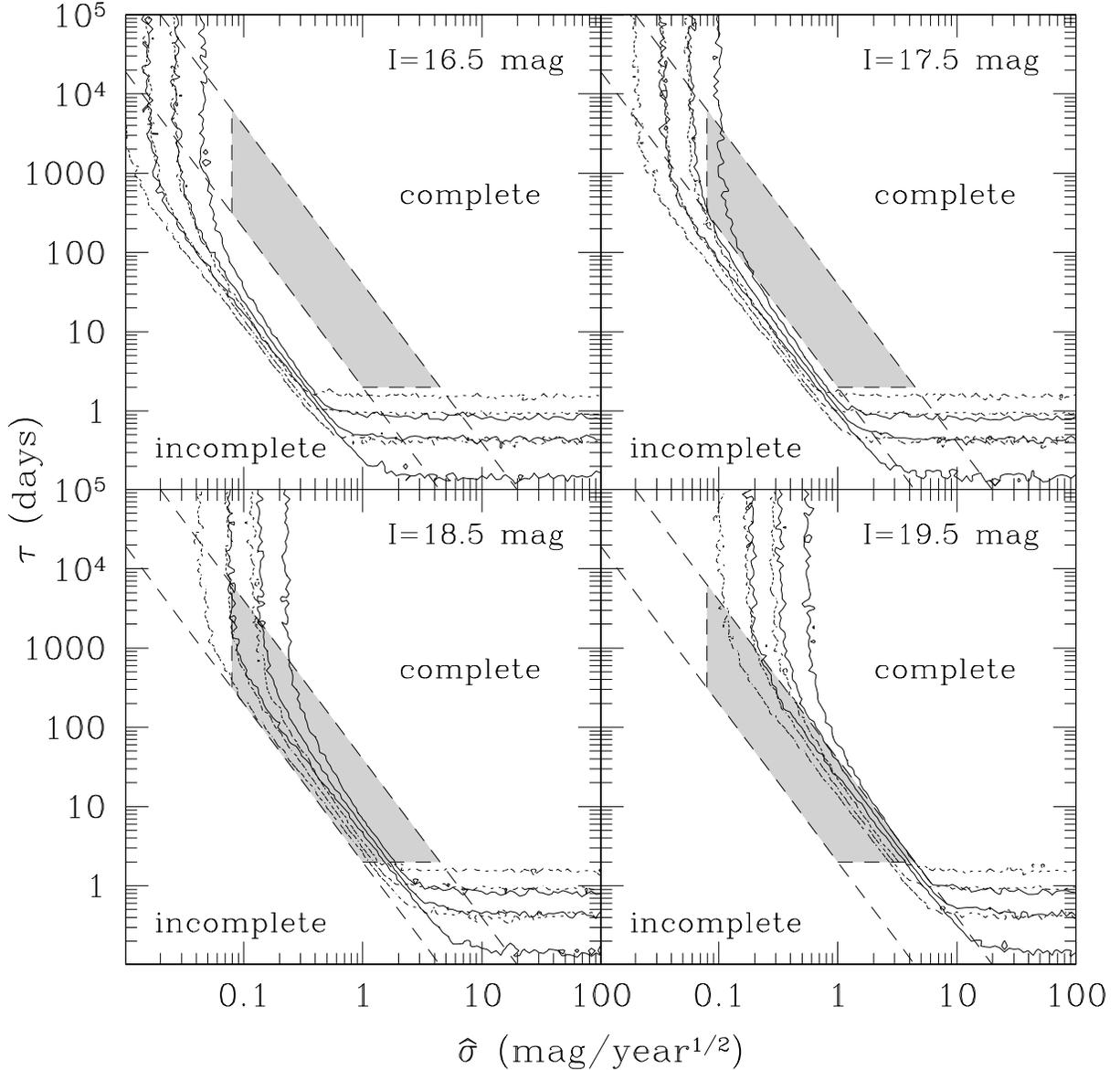}
\caption{
Monte Carlo simulations of completeness for source magnitudes of $\hbox{I}=16.5$ (top left), 17.5 (top right), 18.5 (lower left)
and 19.5~mag (lower right).  The OGLE-II (solid line) and OLGE-III (dotted line) surveys have different cadences, durations and
depths, leading to differences in their completeness limits.
While not plotted, the distributions of variable
sources at a given magnitude closely track these completeness limits
in the sense that variable sources passing the selection criteria are not found 
in the regions where the completeness calculations say they should not be detected.
The three lines show completeness levels of 10\%, 50\%, and 90\% (from left to right). The gray shaded region shows the region
occupied by quasars (Cut 2 of Section~\ref{sec:selection_quasars}).}
\label{fig:oglecomp}
\end{figure}

Figure~\ref{fig:oglecomp} shows the resulting completeness estimates for OGLE-II and OGLE-III
light curves at $I=16.5$, 17.5, 18.5, and, 19.5~mag, where the typical photometric
uncertainties (including the rescalings of Section~2) are $\sigma_{phot}=0.010$, $0.022$, $0.050$, and
$0.12$ mag for OGLE-II and $\sigma_{phot}=0.008$, $0.017$, $0.037$, and $0.087$ mag for OGLE-III.
The general structure of the completeness limit is easily understood.  The variability
criterion can be approximated by noting that in the continuum limit,
where we integrate over a continuous light curve rather examining discreet
samples, the variance of a light curve described by an exponential covariance matrix relative to its mean is
\begin{equation}
  {\rm Var}(c(t)) = \sigma^2 \left[ 1 - { 2 \over x } + { 2 \over x^2 }
      \left(  1 - \exp(-x) \right) \right],
\end{equation}
where $x=\Delta t/\tau$ is the ratio of the survey duration to the time scale $\tau$.
When the survey duration is long compared to the time scale ($x \gg 1$), the signal
variance is simply $\sigma^2$, and light curves above the diagonal line with
$\sigma^2+\sigma_{phot}^2>4\sigma_{phot}^2$ will satisfy the variability criterion.
When the survey is short ($x \ll 1$), the observed variance is limited by the 
survey duration to $\sigma^2 x/3=\hat{\sigma}^2\Delta t/6$ and the completeness
is high if $\hat{\sigma}$ is sufficiently large.  The
criterion that the process is distinguishable from simply increasing the
error bars eliminates sources in the corner with short time scales and
high amplitudes.  Note that the variability criterion is very conservative, because
its selection limits do not improve with additional data.  Using only the
second criterion, based on whether the process model fits better than simply
expanding the measurement errors, would lead to higher completeness for
longer $\tau$ and smaller $\hat{\sigma}$ {\it but at the expense of higher
false positive rates due to systematic errors in the data}.
Compared to the OGLE-II survey, the OGLE-III survey is more sensitive and
covers a longer temporal baseline but with a larger epoch spacing, so it is
more complete for large $\tau$ and smaller $\sigma$, while less
complete for small $\tau$.  The completeness limit is largely
determined by the variability criterion except for the corner with
high amplitudes and short time scales.   While not shown, the distribution
of variable sources as a function of magnitude follows these completeness
limits closely.

\begin{figure}[t]
\centering
\includegraphics[width=16cm]{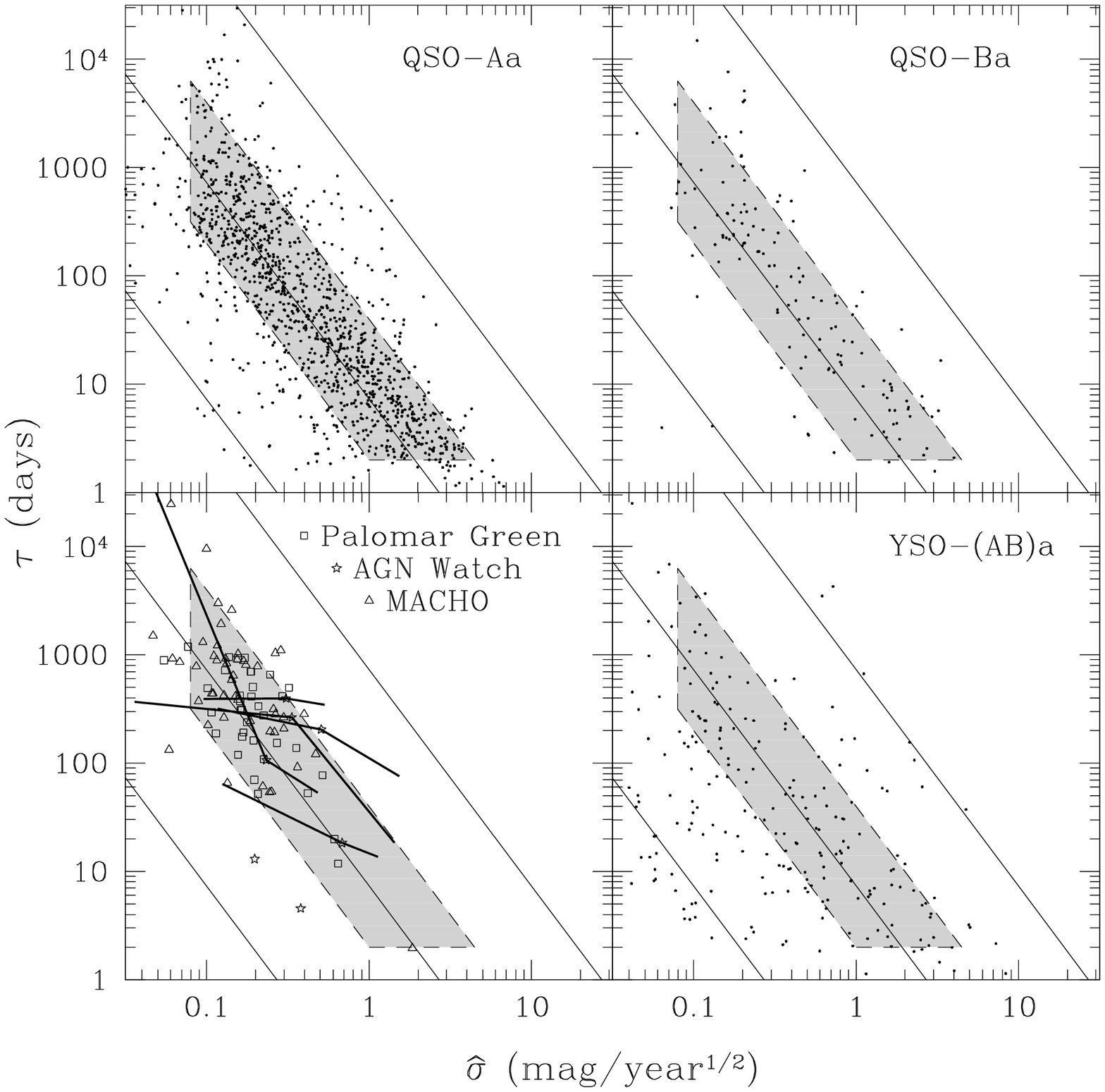}
\caption{Locations of the $\sim 1,000$ variable mid-IR selected quasar candidates from \cite{2009ApJ...701..508K} and $\sim 100$ quasars from \cite{2009ApJ...698..895K} are shown in $\hat{\sigma}$-$\tau$ space. The 
QSO-Aa objects are presented in the top left panel, QSO-Ba in the top right panel, YSO-(AB)a in the bottom right panel, and the \cite{2009ApJ...698..895K} quasars in the bottom left panel. The latter sample is split into Palomar Green (square), AGN Watch (star) and MACHO (triangle) quasars. We also show the lines of constant asymptotic variability $\sigma$ for $0.01, 0.1$, and $1.0$ mag (from left to right) and the region occupied by quasars (Cut 2 of Section~\ref{sec:selection_quasars}, gray area). The thick solid lines in the bottom-left panel are for five AGN Watch quasars, and show the change in the model parameters after adding (subtracting) the host galaxy flux outside (inside) the spectroscopic aperture
used for the light curve. Adding more host moves the source to longer time scales and lower amplitudes.}
\label{fig:sig_tau_type}
\end{figure}

Figure~\ref{fig:qso_intro} shows the distribution of quasars in an optical CMD, 
$\hat{\sigma}_I-\tau$ parameter space (shown in detail in Figure~\ref{fig:sig_tau_type}), $\hat{\sigma}_I-$magnitude space and, where available, the 
relationship between $\hat{\sigma}_I$ and $\hat{\sigma}_V$ at fixed $\tau$. Because the $V$-band light 
curves are poorly sampled compared to the $I$-band, we determine $\hat{\sigma}_V$ using the $\tau$ derived 
for the $I$-band light curve.  
To better understand the contamination problems we discuss later, we superpose a CMD of LMC stars based 
on the {\it HST Local Group Stellar Photometry archive}\footnote{{\tt http://ganymede.nmsu.edu/lg} -- we 
selected two {\it HST} fields {\sc lmc\_u2xq05} and {\sc lmc\_u2xq06}, each observed with the WFPC2 for 1540 and 
1860 seconds in F555V and F814W, respectively. Since in both cases the PC1 chip was pointed at a globular 
cluster, we used the data from the three remaining WF chips, transformed to $V$- and $I$-band. These data were 
then binned into 0.05 mag bins in both magnitude and color and are shown as smoothed contours of 1, 5, 10 
and 20 objects per bin in Figures~\ref{fig:qso_intro}, \ref{fig:variable_intro} and \ref{fig:cmd}.} 
(\citealt{2006ApJS..166..534H}). On the CMD one can quickly identify the red clump (RC), the red giant branch 
(RGB) and the top of the main sequence. Since these are derived from small {\it Hubble Space Telescope (HST)} fields, the sequences of higher 
luminosity stars are not significantly populated.

We see that quasars occupy a well defined locus in the $\hat{\sigma}$-$\tau$ space, lying in a band with
$-2\log_{10}(\hat{\sigma})+0.3 \leq \log_{10}(\tau) \leq -2\log_{10}(\hat{\sigma})+1.6$ that roughly corresponds to 
an asymptotic variability of $\sigma \simeq 0.1$ mag. Variability time scales longer than a few 1000
days are rare, although $\tau$ is indeterminate but long for about 5\% of the sample.  The estimate of
$\tau$ becomes very uncertain as it approaches the survey length. 
At very short $\tau$, systems are lost because they fail the $\ln L_{best} > \ln L_{noise}+2$ criterion. 
If we focus on the QSO-A sources, which are overwhelmingly going to be real quasars, we can define
a shaded region in Figure~\ref{fig:qso_intro} (Cut 2 in Section~\ref{sec:selection_quasars}) that 
encompasses most of these sources (69\% of those identified as variable in OGLE-III).  The more
contaminated QSO-B and YSO candidates are less likely to fall into this region 
(see Figure~\ref{fig:sig_tau_type}, Tables~\ref{tab:results1} and \ref{tab:results2}), providing added
evidence of their higher levels of contamination by LMC sources.  
The mid-IR quasar candidates tend to lie in the Hertzsprung Gap of the CMD, consistent
with the typical color $0.4\leq(V-I)\leq1.0$ mag of $z<3$ quasars.
They also show relatively achromatic variability between
the $V$ and $I$ bands, with more chromaticity for small $\hat{\sigma}$ and larger $\tau$ sources.  
Tables~\ref{tab:results1} and \ref{tab:results2} summarize
the properties of the mid-IR selected quasars for both OGLE-II and OGLE-III light curves.
If we compare the locations of the quasars in the space of the variability parameters
(Figure~\ref{fig:sig_tau_type}) to the completeness estimation of Figure~\ref{fig:oglecomp},
it is easy to understand the magnitude dependence of the fraction of variable quasars in 
Figure~\ref{fig:frac_var}.

The distribution of the quasars is a combination of intrinsic properties (variability at rest wavelength 
$\lambda$ for a black hole of some mass and luminosity), distance (magnitude and scalings between
rest-frame and observed-frame properties), and selection effects (Figure~\ref{fig:oglecomp}).  The
observed magnitude of the quasar depends on distance and {\it K}-corrections, and the variability 
parameters have redshift corrections ($\tau_{rest}=\tau/(1+z)$ and 
$\hat{\sigma}_{rest} = \hat{\sigma}(1+z)^{1/2}$ \citep{2009ApJ...698..895K}) and potentially
{\it K}-corrections for any scalings of the parameters with wavelength, such as those seen in 
average structure functions (e.g., \citealt{2004ApJ...601..692V}).  Our incompleteness depends on observed rather than
intrinsic properties, and this makes it highly likely that our estimate of the region
occupied by quasars is largely correct despite the increasing problems with completeness at
fainter magnitudes.  In essence, the majority of quasars which we lose due to incompleteness
correspond to quasars with similar intrinsic properties at lower redshifts where they
are bright enough to be included, rather than quasars with intrinsically different properties.
The numbers of quasars observed to have a given set 
of parameters are, however, strongly dependent on the selection function, and reconstructing
the true distribution will require redshifts and (essentially) a $V/V_{max}$ method 
applied to the variability selection procedure. A significant advantage of our approach is that
such problems can be quantitatively addressed using the variability model.
The LMC sources we consider in the
next section are an illustration of the opposite limit from the quasars.  Since all
LMC sources are at the same distance, we either detect them in this variability space 
given their magnitude or not.
 
We can also examine the distribution of the \cite{2009ApJ...698..895K} quasars relative
to the mid-infrared sample in Figure~\ref{fig:sig_tau_type}.  They generally lie in the same region (79\% are in the
selection box), but their distribution in that region is very different.  Most of the MACHO and
PG quasars lie at the large $\tau$ end of the distribution, with a few of the 
reverberation mapping quasars dribbling toward short $\tau$.  The MACHO (\citealt{2003AJ....125....1G})
quasars are variability selected, but the selection criterion may be a biased
one compared to a (roughly) magnitude limited sampling of quasars. Figure~\ref{fig:sig_tau_type}
also shows the effects of adding or subtracting host galaxy discussed above.


\section{General Variability of Sources in the LMC}
\label{sec:genvar}

\begin{figure}[t]
\centering
\includegraphics[width=16cm]{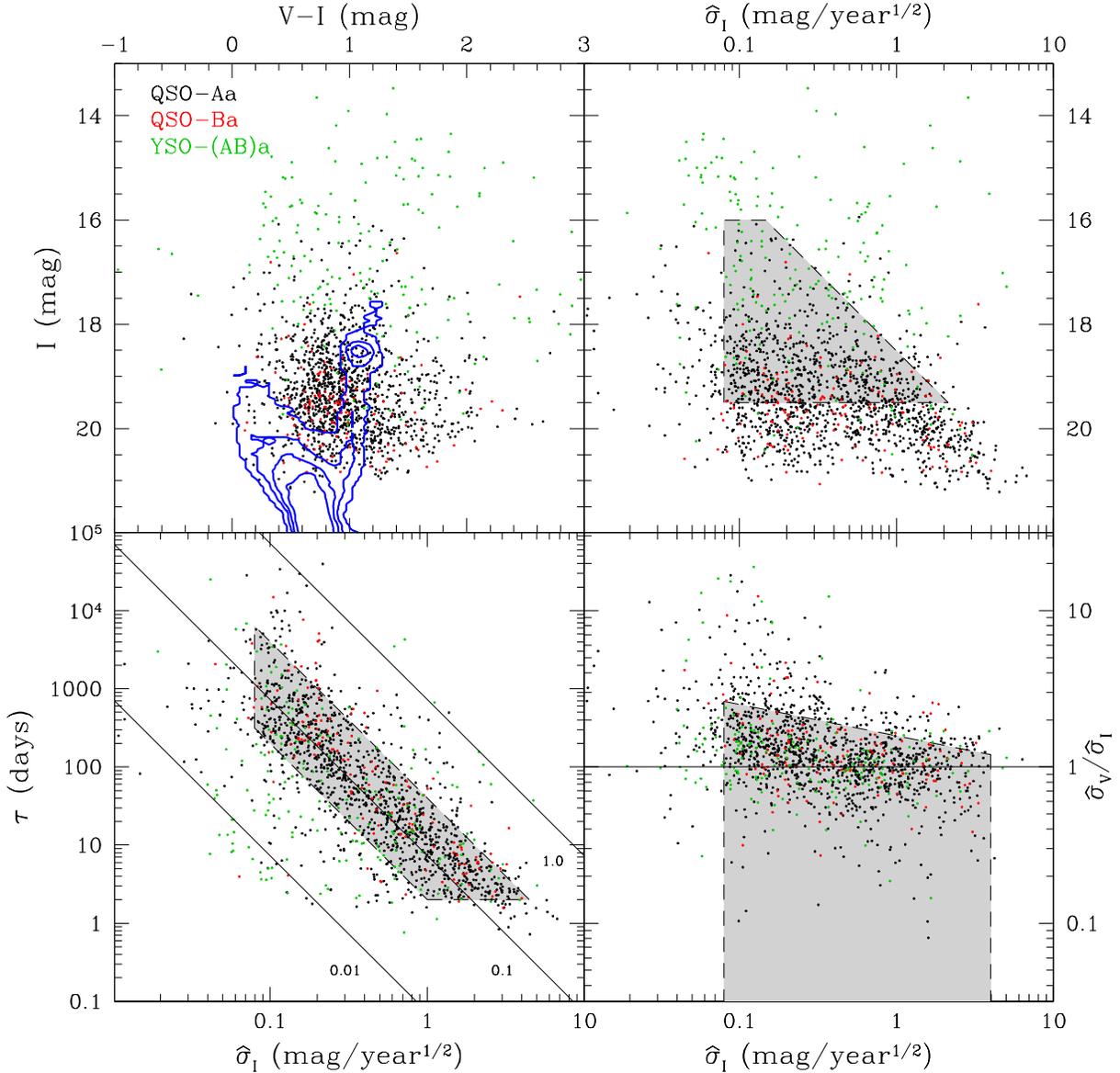}
\caption{Locations of the $\sim 1,000$ variable mid-IR selected quasar candidates from \cite{2009ApJ...701..508K} are shown in the CMD (top left), in $\hat{\sigma}_I$-$I$ space (top right), 
in $\hat{\sigma}$-$\tau$ space (bottom left) and in ($\hat{\sigma}_V/\hat{\sigma}_I$)-$\hat{\sigma}_I$ space (bottom right).
In the top left panel, the contours show the CMD of LMC stars, where the 
contours are for 1, 5, 10, and 20 stars per $0.05\times0.05$ bin, counting from the outer contour. The RC is at $(V-I, I)=(1.05,18.3)$ mag. In the lower left panel we show the lines of constant asymptotic variability $\sigma$ for $0.01, 0.1$, and $1.0$ mag.
The \cite{2009ApJ...701..508K} candidates are coded by color, where the high purity QSO-Aa objects are black, and the high contamination
QSO-Ba and YSO-(AB)a are shown in red and green, respectively. The gray shaded regions indicate the regions occupied by quasars, and are defined by Cuts of Section~\ref{sec:selection_quasars}.}
\label{fig:qso_intro}
\end{figure}

\begin{figure}[t]
\centering
\includegraphics[width=16cm]{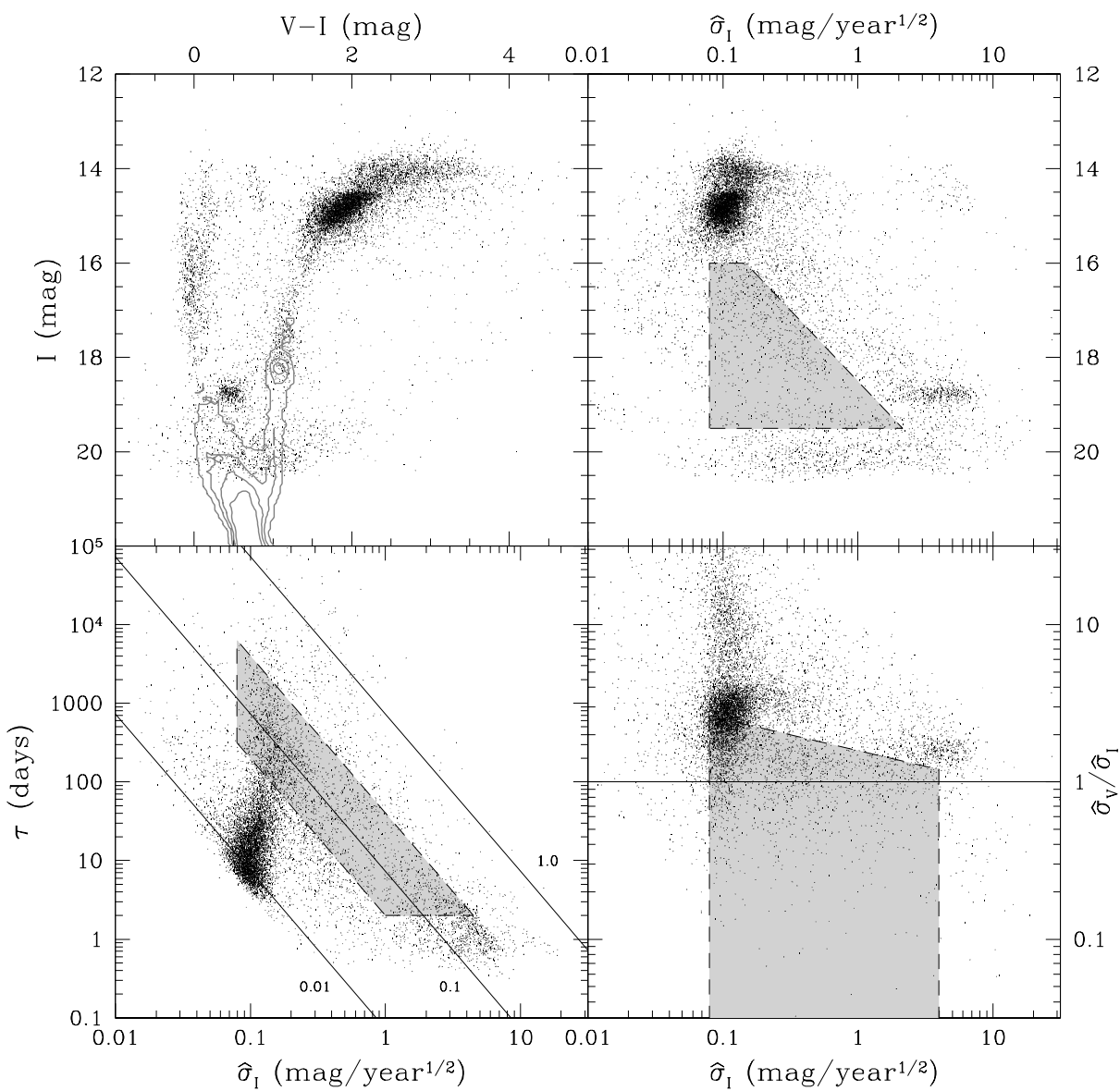}
\caption{Our cleaned sample of $\sim 10,000$ variable objects from the 
six central OGLE-II LMC fields (SC1 -- SC6) satisfying the basic variability criteria from Section~\ref{sec:data}. 
Panels are the same as in Figure~\ref{fig:qso_intro} up to changes in the scales. The shaded regions show the quasar
selection Cuts 2--4 from Section~\ref{sec:selection_quasars}.}
\label{fig:variable_intro}
\end{figure}

In Section~\ref{sec:quasars}, we demonstrated that quasars occupy a limited region of 
the variability parameter space.  In order to select quasars using variability,
we need to examine how other variable sources populate these parameter spaces. 
In Figure~\ref{fig:variable_intro}, we show the location of the general 
variable populations from the inner LMC fields for the same combinations of 
variables used for quasars in Figure~\ref{fig:qso_intro}.
The distributions of outer field variables are similar but of lower density.
Where possible we have identified the variable type using the catalogs discussed in Section~\ref{sec:data}.
Tables~\ref{tab:results1} and \ref{tab:results2} summarize the statistics of selecting sources in these
fields.  It is critical to clean the input catalogs, since the variability
criterion identifies $\sim 86,000$ inner field variables before we apply the bright star masks,
defect masks, and the isolation criteria described in Section~\ref{sec:data}, while
there are only $\sim13,000$ left afterward.  Of these, $\sim 65,000$ (no masks) and 
$\sim 10,000$ (with masks) pass the restriction that the process model fits the data significantly 
better than simply expanding the photometric uncertainties 
($\ln L_{best} > \ln L_{noise}+2$). While the tables show the statistics for both cases,
we only discuss the results for the sources including the masking and isolation criteria.

\begin{figure}[t]
\centering
\includegraphics[width=16cm]{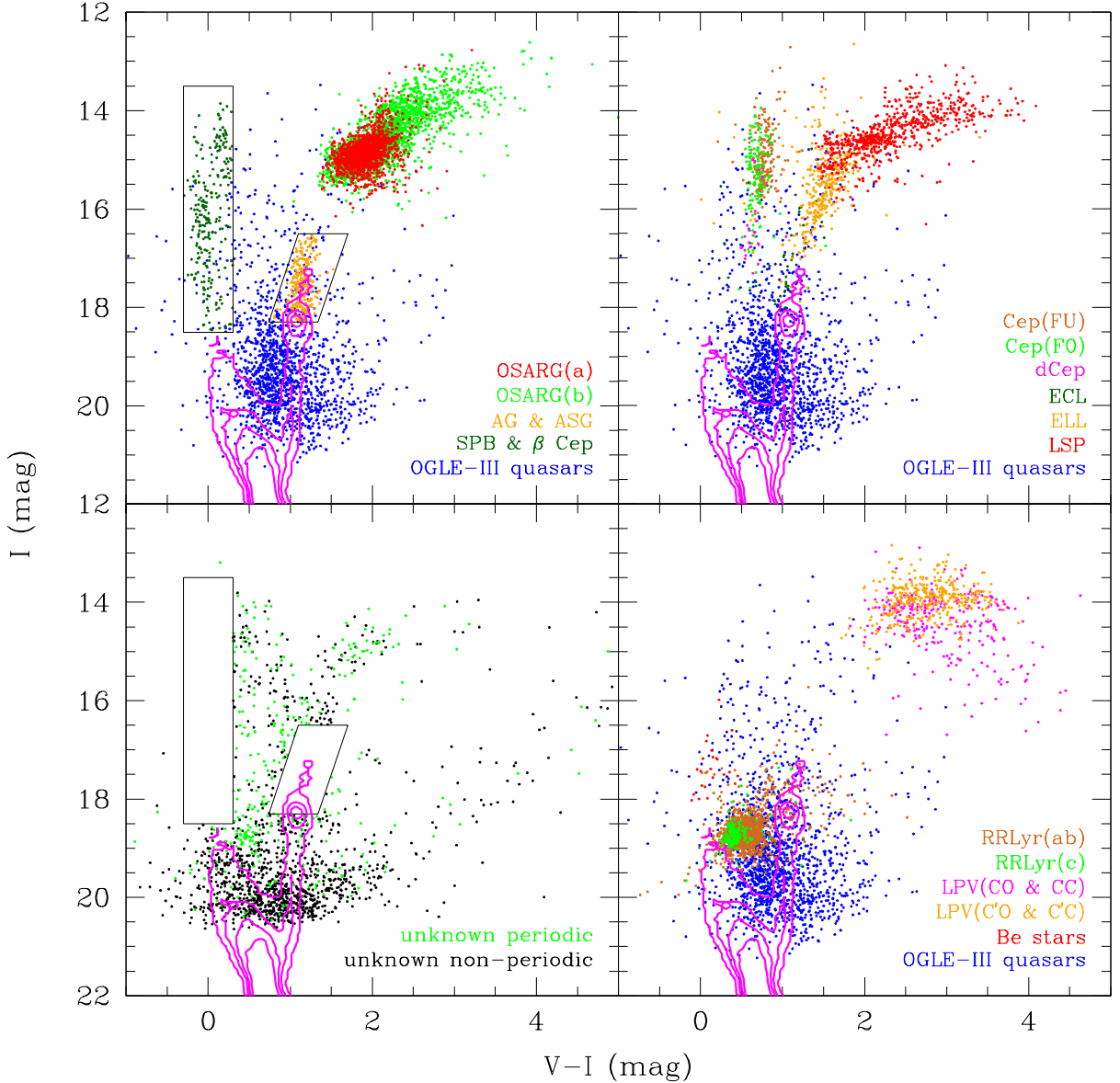}
\caption{CMDs by variable type. In four panels, for clarity, we show 
distinct classes of variable objects. For a detailed description and discussion see text (Section~\ref{sec:genvar}). 
Objects in the two masked areas in the left panels were moved from the bottom panel to the upper one. 
These objects were previously not cataloged by OGLE, but they are known variable classes. In the bottom-left 
panel, we show both periodic and non-periodic objects of unknown origin (or previously overlooked) and in the text 
we discuss the fraction that are quasars. The confirmed quasars and the OGLE-II quasar 
candidates have a similar distribution to the OGLE-III mid-IR selected quasar candidates (blue). For clarity, we 
do not include these smaller quasar samples here or in the subsequent figures.}
\label{fig:cmd}
\end{figure}

\begin{figure}[t]
\centering
\includegraphics[width=16cm]{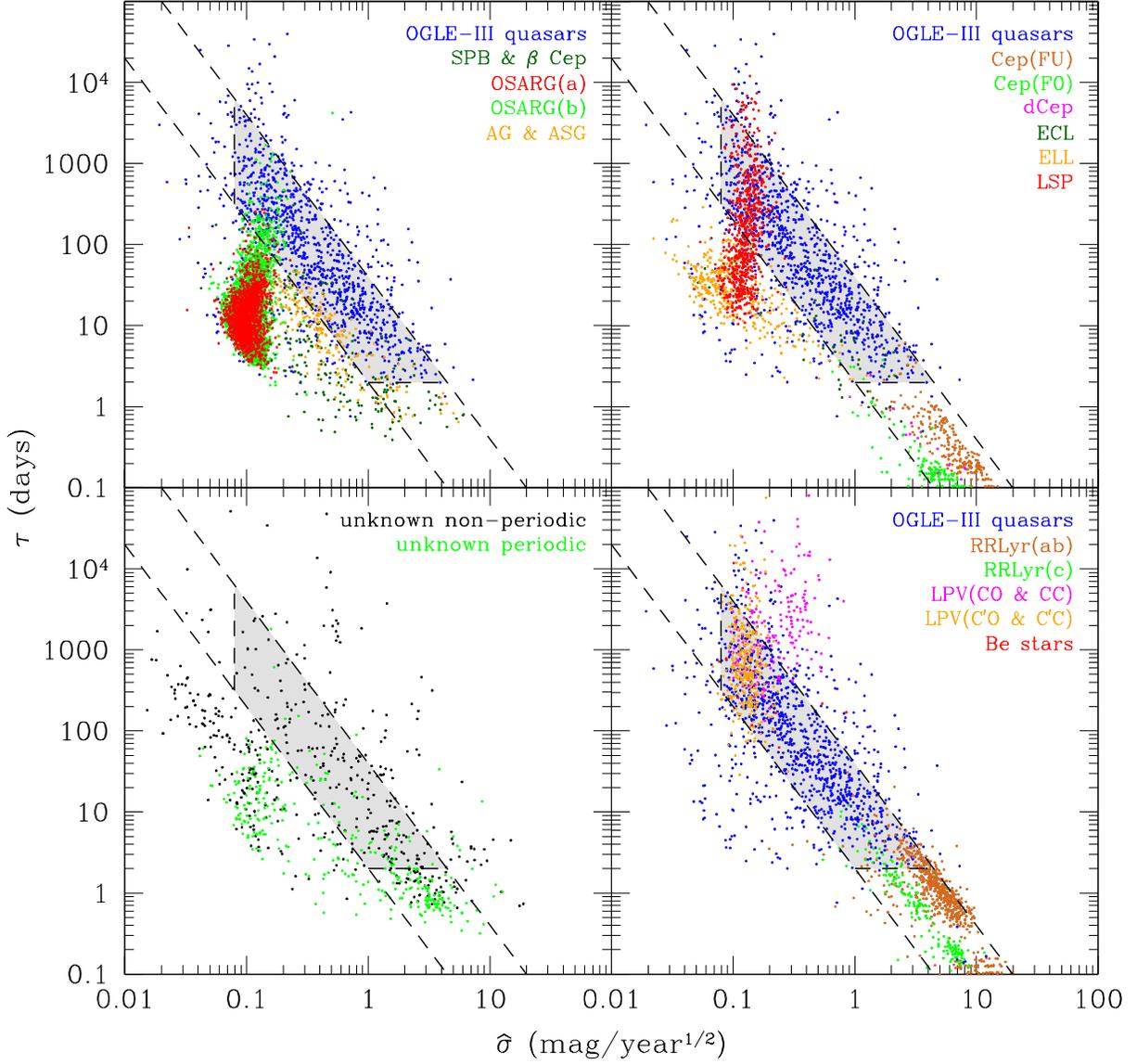}
\caption{Characteristic time as a function of modified amplitude ($\hat{\sigma}$-$\tau$). The four panels show the same 
variability classes as in Figure~\ref{fig:cmd}. The gray band is the quasar locus (Cut 2 in Section~\ref{sec:selection_quasars}) 
used to separate quasars from the unknown non-periodic objects shown in the bottom-left panel.}
\label{fig:sig-tau}
\end{figure}

\begin{figure}[t]
\centering
\includegraphics[width=16cm]{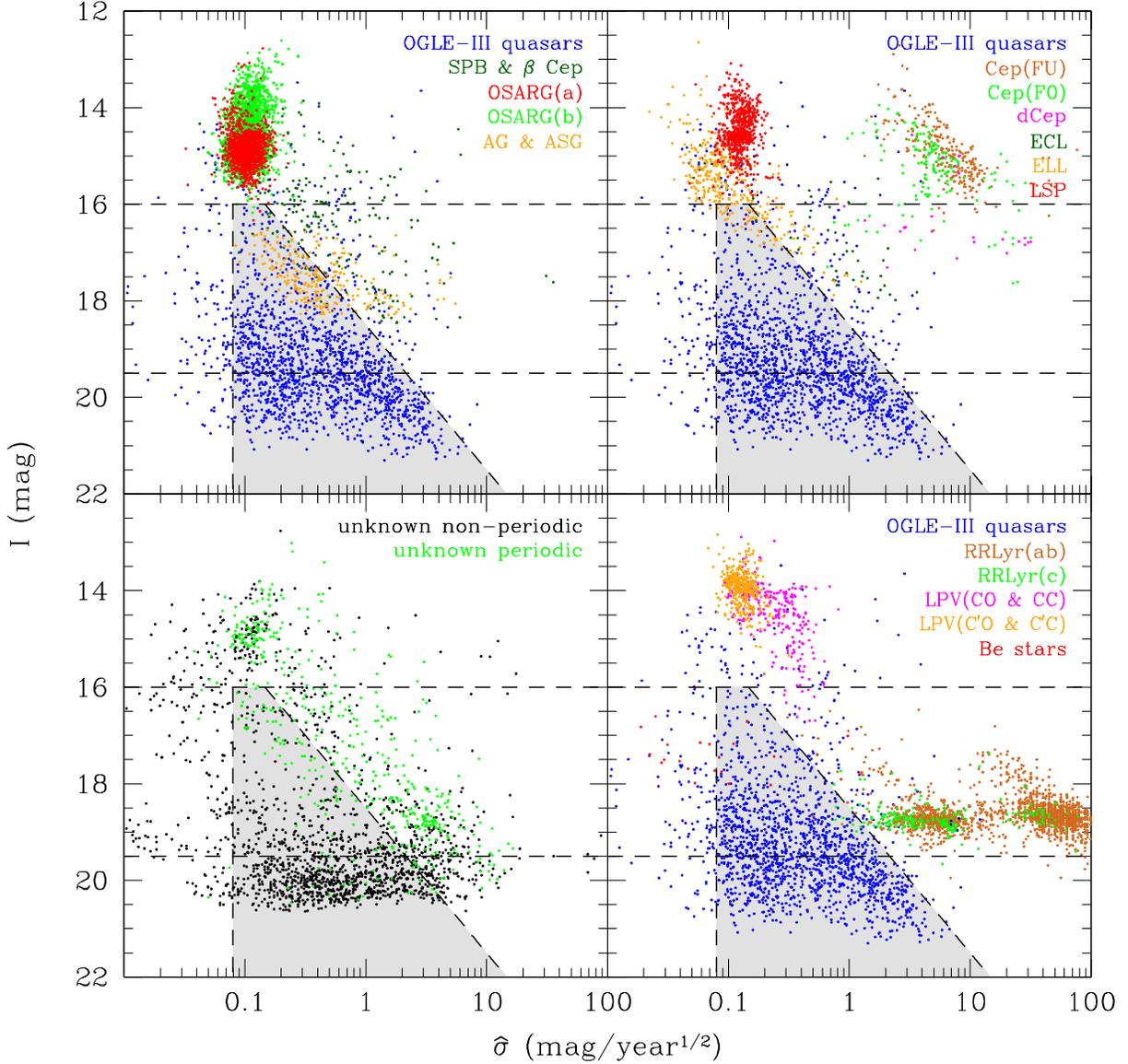} 
\caption{Modified amplitude vs. magnitude ($\hat{\sigma}$-I). The four panels show the same variability classes as in Figure~\ref{fig:cmd}. The gray area is the region defined by a high density of OGLE-III quasars, which separates them from other variability classes (Cut 3 in Section~\ref{sec:selection_quasars}). The heaviest contamination comes from the active giants and subgiants (AG/ASG, top left), and hot blue pulsators (SPB/$\beta$ Cep, top left) but these can be ruled out based on their periods (see Figure~\ref{fig:tau_p}). The gray wedge can in principle 
extend to arbitrarily faint magnitudes, but must be restricted based on the survey depth. For example, OGLE-II objects fainter than $I>19.5$ mag are mostly false variables (below bottom horizontal dashed line). }
\label{fig:sig-mag}
\end{figure}

\begin{figure}[t]
\centering
\includegraphics[width=16cm]{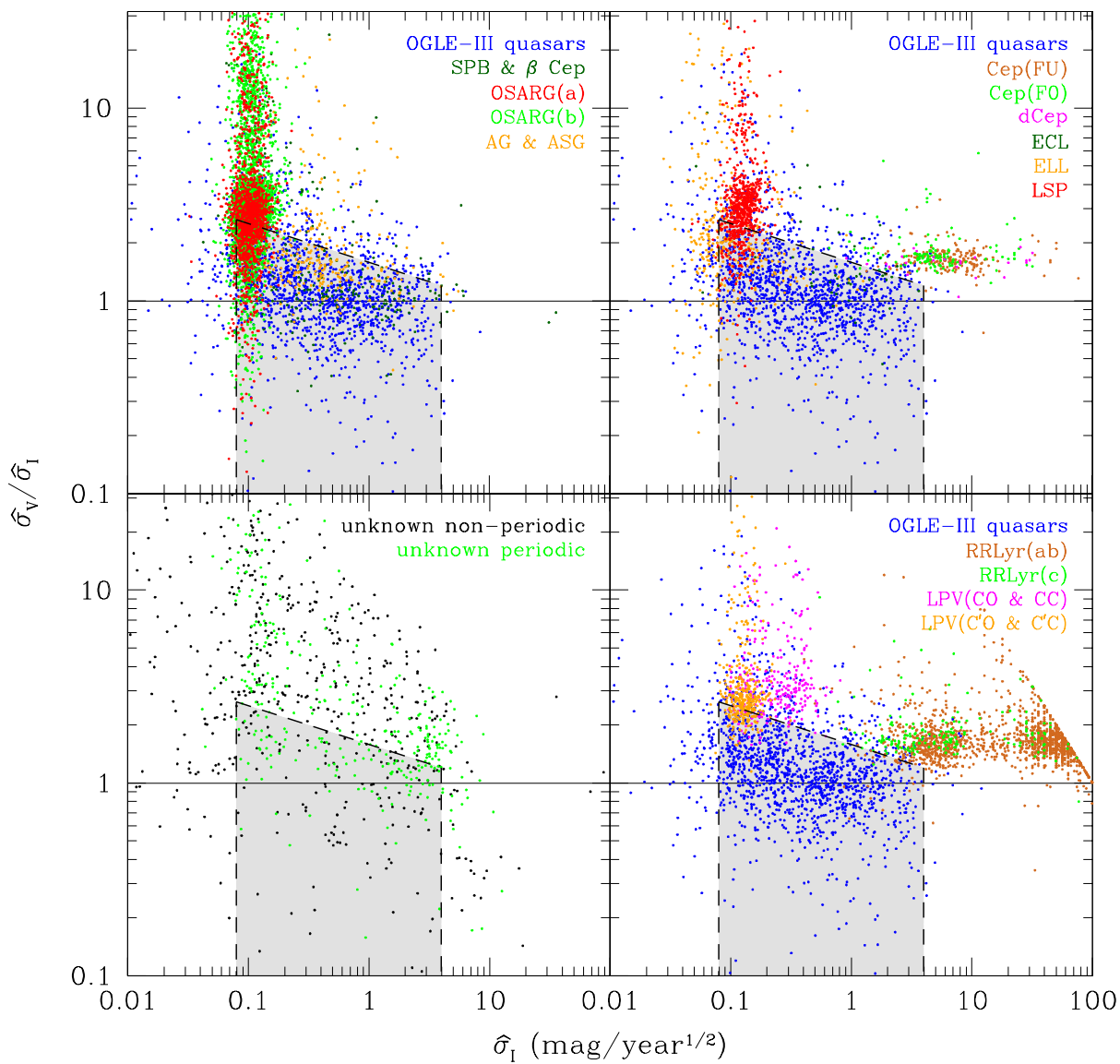} 
\caption{Ratio of variability amplitudes in the $V$- and $I$-bands. It is clear that stellar objects generally vary more in the $V$-band
 than in the $I$-band, while QSOs show smaller differences. The gray area is the region described by Cut 4 of Section~\ref{sec:selection_quasars}.}
\label{fig:sig-VI}
\end{figure}

\begin{figure}[t]
\centering
\includegraphics[width=16cm]{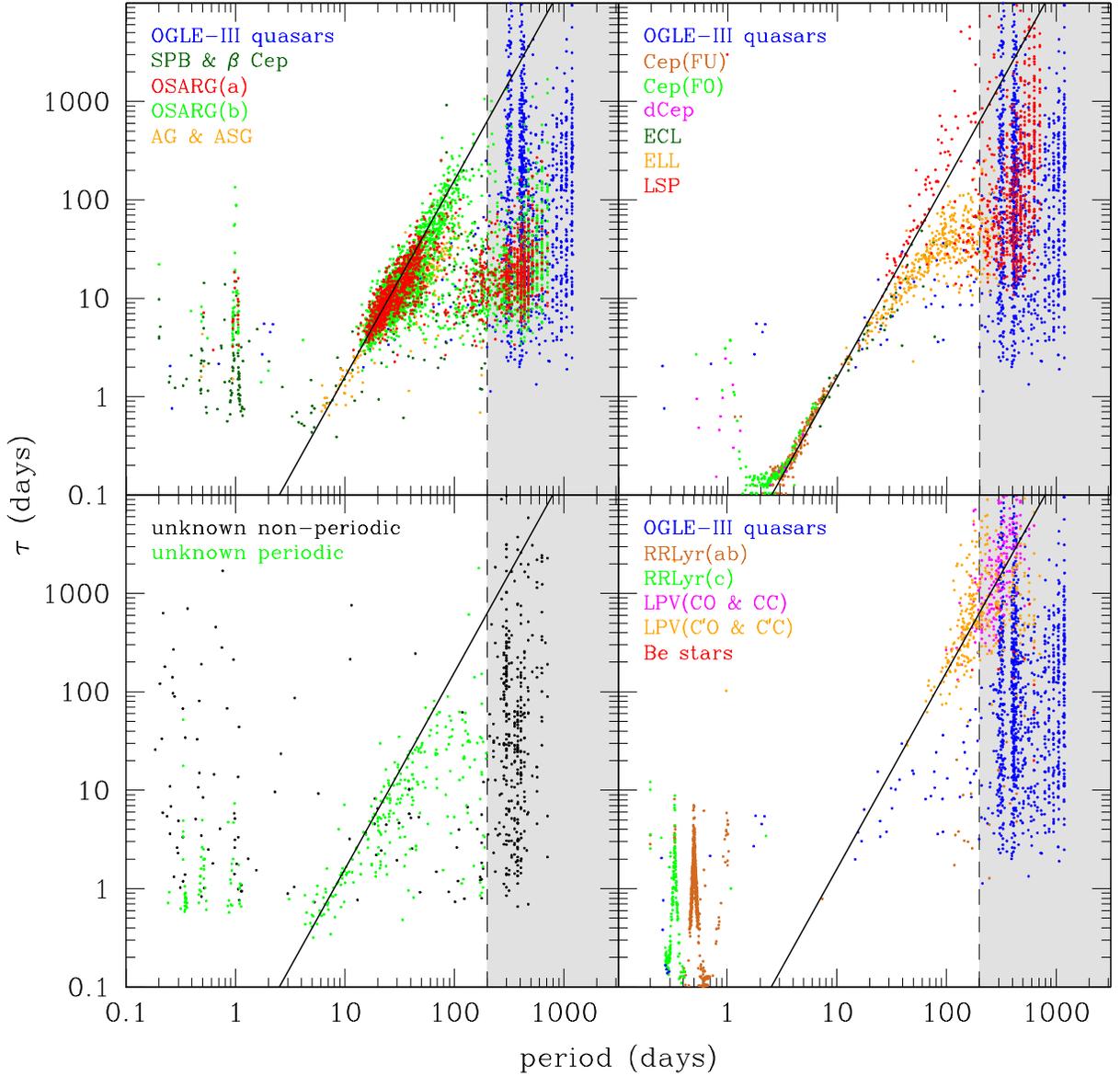} 
\caption{Relation between Fourier periods $P$ and time-scales $\tau$. These are not fundamental and depend on the light curve
sampling, particularly on the mean epoch spacing.  Except for the bottom left panel, the panels show 
objects with a high probability of being periodic ($\log_{10}(p_{\rm periodic}) < -3$) and non-periodic quasar candidates (blue dots, here 
selected with $\log_{10}(p_{\rm periodic}) > -3$). Many periodic variable objects show a clear correlation of the characteristic 
time-scale $\tau$ with period $P$. These are the Cepheids, OSARGs and long period variables.
Quasars clearly show no such dependence. The bifurcation in period for periodic objects at $\tau\geq20$ days (top panels) is 
probably due to the multi-mode pulsations of OSARGs and LSPs. 
The solid line corresponds to a $\tau$-period relation described by $\log_{10}(\tau)=2\log_{10}(P)-1.8.$ The objects in the bottom left 
panel are previously unrecognized or overlooked non-periodic objects (black, $\log_{10}(p_{\rm periodic}) > -3$ or $P>200$ days) and 
periodic variables (green, $\log_{10}(p_{\rm periodic}) < -3$ and $P<200$ days). Spikes at 1, 1/2, 1/3 and 1/4 days are due to aliasing 
problems in the periodograms that are not fully excluded by the narrow period notches we use to suppress
most diurnal aliasing problems.}
\label{fig:tau_p}
\end{figure}

In Figure~\ref{fig:variable_intro}, we can quickly identify a number of 
sequences or overdensities of variable objects. We separate them, for clarity, in the four panels of 
Figures~\ref{fig:cmd}--\ref{fig:tau_p}. In these figures we plot rare samples last
so that they are not masked by the common ones. Quasars are shown in multiple panels
to maximize their visibility compared to other populations, and Figures~\ref{fig:qso_intro} and \ref{fig:sig_tau_type} are always
available for reference. In the top-left panel of each figure we show two samples of 
OSARGs, where ``a'' corresponds to the asymptotic giant branch (AGB) objects and ``b'' to the RGB 
objects (\citealt{2007AcA....57..201S}). They are approximately 3 mag brighter than the RC 
and also redder. In this panel, we also show slowly pulsating B stars (SPB, \citealt{2006MmSAI..77..336K}), 
active giants, active subgiants, and  RS CVn binaries (AG and ASG, \citealt{2006AJ....131.1044D}). 
There is no formal OGLE catalog of these variables, so we defined them based on the selection boxes 
shown in Figure~\ref{fig:cmd}. The OSARGs share the same area of the CMD with the LSPs, ELLs (top-right panel),
and LPVs (bottom-right panel). The OGLE eclipsing binary samples are located just above the RC, and, slightly 
higher, one finds ellipsoidal variable red giants (top right). Along the instability strip with an 
approximate color of $(V-I)=0.5$, one can find Cepheids, $\sim 3$ mag brighter than RC, and RR Lyrae, 
which are slightly fainter then RCGs. The fundamental-mode Cepheids (RR Lyrae), denoted as ``FU'' (``ab''), 
and first-overtone ones, denoted as ``FO'' (``c''), form distinct sequences. \cite{2003AJ....125....1G} in their 
attempt to select QSOs based on variability found that Be/Ae stars are the main source of contamination. 
Be stars can be found in the bottom-right panel just above the main sequence. In the bottom-left panel we 
present all the remaining variable objects. These should be quasars, other unclassified variables and 
sources with unidentified systematic problems.  For comparison, we also show the OGLE-III quasar candidates 
in three of the panels, excluding the panel of unclassified sources.

The top-right panels of Figures~\ref{fig:qso_intro} and \ref{fig:variable_intro} show the positions of objects in 
the ``modified amplitude''--magnitude ($\hat{\sigma}-$I) space.
Once again, a number of source groupings are easily noticed, where the most prominent ones are OSARGs, LSPs, LPVs, RR Lyrae, and Cepheids
(color coded in Figure~\ref{fig:sig-mag}). OSARGs, ELLs, LSPs, and LPVs overlap, although they can be partially distinguished based on their periods (Figure~\ref{fig:tau_p}). In $\hat{\sigma}-$I space Cepheids and RR Lyrae occupy distinct areas.

We show the positions of all the variable stars and quasars in the space of the stochastic model parameters in Figures~\ref{fig:qso_intro} 
and \ref{fig:variable_intro} (bottom-left panel). In this space of modified ``characteristic time'' and 
``variability amplitude'' ($\tau-\hat{\sigma}$), we color coded all variability classes and present them in
Figure~\ref{fig:sig-tau}. A majority of OSARGs and a substantial fraction of ELLs, 
LSPs, and LPVs can be easily distinguished from the quasar locus, although there is heavy contamination for long $\tau$. 
Note also that this is also the region occupied by most MACHO quasars, and this may be one cause of low yield 
of variability selected quasars in Geha et al. (2003).
Cepheids, RR Lyrae, some LPVs, Be stars, ECLs, SPBs, and $\beta$ Cep, AGs and ASGs also overlap the
quasar locus area. Note that fundamental mode and first overtone Cepheids
and RR Lyrae can be distinguished from each other in $\hat{\sigma}-\tau$ space.

The bottom-right panels of Figures~\ref{fig:qso_intro} and \ref{fig:variable_intro} show the variability ratio 
between the $V$- and $I$-bands as a function of $I$-band variability. The majority of LMC objects show higher 
variability amplitudes in the $V$-band than in the $I$-band, as we would expect since stellar variability is dominated
by temperature changes.  Quasar variability is less chromatic, so the amplitude ratio is another tool for
separating quasars from variable stars.  For these LMC fields, this is not critical because of the large
magnitude differences between many of the variable stars and quasars.  However, in a field dominated by Galactic
sources spread over a wide range of distances, amplitude ratios have the advantage of being distance
independent.  The color-coded variability classes are shown in Figure~\ref{fig:sig-VI}. 

Figure \ref{fig:tau_p} shows all the variable sources and quasars (blue dots) in the space of the 
characteristic time-scale $\tau$ and the most likely period $P$ derived from a standard periodogram.  
As noted in Section~\ref{sec:method}, quasars never have significant probabilities of being periodic
sources unless the ``period'' is long, $P \gtorder 200$~days.  On inspection, these light curves have fluctuations on
long time scales that are quasi-periodic over the extent of the light curve (particularly with
yearly or half-yearly aliases), so a sine wave
fit to the data is a significantly better fit to the data than a constant.  Even the quasars
with falsely probable periods show no correlation between the period and $\tau$. This leads us to 
define a periodic source as one with $P<200$~days and $\log_{10}(p_{periodic}) <-3$.  On short
time scales ($\tau$ or $P \ltorder 1$~day) we see no real correlations because the periodograms
have diurnal aliasing problems and many sources fail to pass the  $\ln L_{best} > \ln L_{noise}+2$
variability selection criterion.  For periods of days to tens of days, many sources follow
a scaling with $\tau \propto P^2$, which is surprising since we expect $\tau \propto P$ by
dimensional analysis. 

Our one hypothesis for this scaling is that it can be produced if 
the exponential process model is dominated by how well it fits the peak of the correlation
function. Near the peak, the auto-correlation function of a sine wave is proportional to 
$\Delta t^2/P^2 +const$, while that of the exponential covariance matrix is $\Delta t/\tau + const$,
potentially driving the $\tau \propto P^2$ scaling.  There is a regime where the longer
period Cepheids have $\tau \propto P$, and some of the multiperiodic sources (OSARGs)
appear in multiple groupings. Monte Carlo experiments fitting sine wave light curves show
that the relation between $\tau$ and $P$ is strongly affected by the temporal sampling, 
with the $\tau \propto P^2$ region shifting to longer periods as the mean epoch spacing
increases. Despite some effort we could not derive an analytical model for the scaling.
It is clear, however, that the relationship between $\tau$ and $P$ depends on sampling 
rather than being universal.


\section{Variability Selection of Quasars}
\label{sec:selection_quasars}

We can now combine the results from Sections~\ref{sec:quasars} and \ref{sec:genvar} to define 
selection criteria for isolating quasars from other variable sources.  Presently this is 
limited to searching for candidate quasars in the 1.3~deg$^2$ (0.65~deg$^2$) covered by the OGLE-II inner (outer) fields
in order to explore how well we can find quasars while avoiding contamination. We
limit the analysis to sources with $I<19.5$ based on the completeness estimates
in Section~\ref{sec:quasars}.  There should be very few quasars in this region, so this is
really a test of contamination levels in a field with very high densities of stellar 
variables.  Based on \cite{2006AJ....131.2766R} we expect approximately 1, 6, and 20
quasars per square degree with $I<17.5$, $18.5$, and $19.5$~mag respectively,
which agrees well with the statistics of the mid-IR selected quasars (\citealt{2009ApJ...701..508K}).  
We will simultaneously apply the selection method to the broader sample
of OGLE-II and OGLE-III quasar light curves to examine their effects on completeness.
We select quasars in four steps, starting with two that will be generic
to any set of light curves from any location (Cut 1 -- non-periodicity, Cut 2 -- variability properties of quasars),
and then applying those that depend on the specific field (Cut 3) and the availability
of light curves in multiple bands (Cut 4).
We will not apply a color selection criterion, since one goal of any wide application
of this method would be to find quasars at redshifts where their colors are
crossing the stellar color distribution near $z\sim 2.6$ (\citealt{2006AJ....131.2766R}).

{\it Cut 1.} We start by eliminating periodic sources. A periodic source is defined to have a maximum likelihood
   period $P<200$~days with $\ln (p_{\rm periodic}) > -3$.  This removes 76\% of the sources,
   including, 99\% of RR Lyrae, 100\% of Cepheids, 91\% of ELLs, and 100\% of ECLs. It
   does not remove 20\% of OSARGs, 69\% of LSPs, and 36\% of LPVs. 

{\it Cut 2.} Next we isolate sources with the variability properties of quasars, defined
  by the region in $\hat{\sigma}-\tau$ space bounded by 
  $-2\log_{10}(\hat{\sigma})+0.3<\log_{10}(\tau)<-2\log_{10}(\hat{\sigma})+1.6$, $\log_{10}(\hat{\sigma})>-1.1$ 
  and $\tau>2$ days.  This quasar region is shown in gray in Figures~\ref{fig:oglecomp}, \ref{fig:qso_intro}, 
  \ref{fig:sig_tau_type}, \ref{fig:variable_intro} and \ref{fig:sig-tau}.
  The lower limits are designed to remove most Cepheids, RR Lyrae, OSARGs, ellipsoidal variable 
  red giants, long secondary period variables, slowly pulsating B stars, $\beta$ Cephei stars, 
  active giants and subgiants, and YSO-(AB) objects.  The upper limits remove many of the
  long period and long secondary period variables.  This criterion will still leave 
  significant contamination from several of the variable star populations along the
  left (low $\hat{\sigma}$) and bottom edges. These are blue variables, active giants and subgiants and RR Lyrae.
  This cut removes 97\% of ELLs, 92\% of OSARGs, 45\% of LSPs, and 41\% of LPVs passing Cut 1.

{\it Cut 3.} The third cut takes advantage of the fact that almost all the contaminating sources
  are at a common distance and largely consist of relatively bright giant stars.  It is defined
  by a wedge in $\hat{\sigma}-$magnitude space that isolates the OGLE-III quasars.  
  We first require quasar candidates to be fainter than $I=16$~mag since such
  bright quasars are very rare, while the OSARGs, classical Cepheids, eclipsing binaries, 
  ellipsoidal variable red giants and long period variables are all brighter than $I = 16$ 
  mag given the distance to the LMC. A second limit, $I > 3 \log_{10}(\hat{\sigma}) + 18.5$, 
  separates the RR Lyrae, double mode Cepheids, ECLs, some ELLs, SPBs, $\beta$ Cep, AGs and 
  ASGs, which are to the right of the quasar region. As in Cut 1, we use the same lower limit 
  on the scaled amplitude of $\log_{10}(\hat{\sigma}) >-1.1$, which partially eliminates Be stars. 
  This cut removes 100\% of the ELLs, LPVs, LSPs and OSARGs passing Cut 2.
  Cut 3 would need to be adjusted for other background environments, such
  as the Galactic bulge, where similar variable stars
  are $\sim4$ mag brighter than in the LMC (depending on the amount of extinction).

{\it Cut 4.} For the objects with well determined $V$-band amplitudes $\hat{\sigma}_V$ we can
  also separate quasars using the ratio of the $V$- and $I$-band amplitudes. 
  We define an area in $(\hat{\sigma}_V/\hat{\sigma}_I)$-$\hat{\sigma}_I$ space with 
  $-1.1<\log_{10}(\hat{\sigma}_I)<0.6$ and $\log_{10}(\hat{\sigma}_V/\hat{\sigma}_I)<-0.2\log_{10}(\hat{\sigma}_I)+0.2$, 
  where we select objects with a smaller ratio of the $V$- to $I$-band variability amplitudes 
  than is typical for stars (Figure~\ref{fig:sig-VI}, gray area).
  This cut removes most Cepheids and RR Lyrae and a substantial fraction of other common variable stars.

The effects of the selection functions on both the LMC variable population and these
quasar samples are summarized in Tables~\ref{tab:results1} and \ref{tab:results2} both for their effects
when applied sequentially (Table~\ref{tab:results1}) and individually (Table~\ref{tab:results2}).  We also summarize the effects on the
sub-categories of the mid-IR selected samples, where we expect to eliminate more
of the heavily contaminated QSO-B and YSO classes than the relatively high 
purity QSO-A class.  Recall from Section~\ref{sec:quasars} that we lose a large fraction
of the quasars from the two requirements for being considered a variable because most 
quasars are relatively faint compared to the depths of the OGLE data. 

The first test of a selection method is its completeness. We evaluate this based on the
on the $\sim 2,700$ ($\sim 300$) mid-IR 
selected quasar candidates from \cite{2009ApJ...701..508K} with OGLE-III (OGLE-II) light curves
as well as the smaller \cite{2009ApJ...698..895K} sample.  The biggest 
problem for variability selected quasars from these samples is that most quasars will be faint. While we can
identify 2716 (310) of our mid-IR quasars in the OGLE-III (OGLE-II) data, only 1139 (215) are bright enough ($I<19.5$ mag)
to plausibly detect quasar variability (see Figure~\ref{fig:frac_var}).  For the full 
OGLE-III (OGLE-II) samples, 984 (58) pass the basic variability criterion of Section~\ref{sec:data},
while 635 (58) of the $I<19.5$ mag candidates pass this criterion.  Adding the noise
criterion, $\ln L_{best} > \ln L_{noise}+2$, we find 619 (848) for OGLE-III and 55 (55) for OGLE-II, 
brighter than  $I<19.5$ mag (no magnitude limit).  The difference in the fraction of variable objects 
in the two phases of the OGLE survey is explained by the added depth (0.3 mag)
and duration (4 versus 8 years) of the OGLE-III survey.  By design, the remaining cuts have
little effect on the quasar candidates, although we lose a larger fraction of the non-QSO-A
sub-samples, as we would expect given their higher levels of contamination. After applying the
first three cuts to the variable objects, we are left with 65\% of the QSO-A sample.
We can apply only Cuts 1 and 2 to the \cite{2009ApJ...698..895K} sample, and 
65\% of the sources pass the cuts. Cut 4 removes an additional 7\% of the OGLE-III QSO-Aa objects. 
Finally, all four cuts are met by 61\% (63\%) of the OGLE-III QSO-Aa quasar candidates
with no faint magnitude limit (within $16<I<19.5$ mag) and 17\% of the OGLE-II candidates.
Thus, following this approach can yield variability-selected samples of quasars with moderate completeness. 

Next we examine how well we reject other variable sources, focusing on our analysis of the
central region of the LMC.  We only discuss the sample including masks and isolation 
criteria (see Section~\ref{sec:data}), but report the results for the full, unpruned sample as well 
in Tables~\ref{tab:results1} and \ref{tab:results2}.
The selection cuts are very effective at reducing the numbers of candidate sources, going from $\sim10,000$ ($\sim600$)
sources passing the variability criterion in the inner (outer) fields to only 58 (2) candidate quasars.  In the outer fields with
fewer stars, the contamination is remarkably low.

Of the 58 remaining OGLE-II candidates in the inner fields, 3 are also mid-IR selected quasars.
There are 63 (41 QSO-A) OGLE-II sources in the same fields, 24 (12) passing the two initial variability criteria,
showing that the biggest problem for quasars is recognizing variability given the depth of the data. 
The mid-IR quasar selection method is not perfect, in particular
it tends to miss lower redshift quasars with significant mid-IR emission from the host galaxy
(\citealt{2008ApJ...679.1040G,2009arXiv0909.3849A}), so some of the
remaining, unidentified candidates may be real quasars.   For the \cite{2006AJ....131.2766R}
quasar number counts, we expect of order $1$, $8$, and $25$ quasars brighter than $I<17.5$,
$18.5$ and, $19.5$~mag in this 1.3 deg$^2$ region, compared to 6, 16 and 41 QSO-Aa candidates in OGLE-II.   
We examined the light curves of all 58 objects that passed the four criteria. 
On visual inspection, the majority seem to be created by photometric problems in the wings of other stars,
long period variable stars just above the RCG, stars in the wings of high proper motion stars (see \citealt{2001MNRAS.327..601E}),
blue and Be stars, eruptive variables, novae/supernovae,
nova-like variable, unidentified types of variable stars, and quasar candidates. 
Of the non-mid-IR sources, only one seems likely to be a quasar, the OGLE-II (OGLE-III) object LMC\_SC5 253536 (LMC167.2 36755, at J2000 05:24:24.17 --69:55:56.9). It has a brightness of $I=18.53$ (18.37) mag in OGLE-II (OGLE-III).
Its color of $(V-I)=1.38$ (1.06) mag is somewhat redder than expected for $z<3$ quasars. 
We show both the OGLE-II and OGLE-III light curves in Figure~\ref{fig:lc}. 

In the outer OGLE-II fields we are left with only 2 candidates. On inspection we find no plausible QSO candidates.
None of the OGLE-II light curves of the 43 mid-IR selected candidates (29 QSO-Aa) from the outer fields passes all 
the criteria. In fact, only two QSO-Aa and two YSO-Aa objects with OGLE-II light curves pass the two initial variability criteria. 

\section{Summary}
\label{sec:summary}

Periodograms to recognize periodically varying sources and begin their classification are
a standard astronomical tool.  Here we introduce an equally simple approach to
classifying continuously varying sources that are not periodic and explored the properties
of large samples of variable sources in the LMC and SMC using data from the OGLE-II and OGLE-III
microlensing projects. We modeled the sources as a damped random walk, a stochastic process with an
exponential covariance matrix characterized by a scaled amplitude $\hat{\sigma}$, a time
scale $\tau$, and an asymptotic variance on long time scales of $\sigma = \hat{\sigma}(\tau/2)^{1/2}$. 
This model was introduced by  \cite{2009ApJ...698..895K}, who used a forecasting approach
to estimate the parameters that is not as statistically powerful as the PRH method used
here.  Extracting the damped random walk parameters is no more complex
than obtaining a periodogram, so it can be easily used to characterize any light curve.

We first applied this method to a much larger sample of quasars than \cite{2009ApJ...698..895K}
by using the mid-IR selected quasars from \cite{2009ApJ...701..508K} that also have OGLE-II
or OGLE-III light curves.  We have verified the suggestion by \cite{2009ApJ...698..895K}
that quasar light curves are well modeled by the damped random walk process.  Moreover,
we find that quasars occupy a well defined region of the parameter space of the model,
allowing us to design a simple quantitative method for the variability selection of quasars.
The details of the approach will likely require some modifications as larger samples of
quasars are analyzed and the method is applied in regions with different sources of
stellar contamination, but the broad outline should survive. We also note that the model
allows quantitative estimation of completeness in variability selected quasar samples
and can estimate the detection ranges needed for computing luminosity functions
or other population statistics.

While \cite{2009ApJ...698..895K} are certainly correct that a damped random walk 
corresponding to an exponential covariance matrix models quasar variability well,
it is less clear that the correlations between the model parameters and the 
luminosity or black hole masses of quasars are robust.  Since we lack redshifts
for our quasar sample we cannot explore this in detail, but we note two  
potential systematic problems in the  \cite{2009ApJ...698..895K} analysis.
First, the distribution of the \cite{2009ApJ...698..895K} quasar sample 
may be a biased sampling of the quasar distribution. It
appears to be biased toward sources with long time variability time scales compared
to the mid-IR quasar sample.  Second, most of the leverage for determining the slopes
in the relations between $\tau$ and $\hat{\sigma}$ and the quasar mass and
luminosity in \cite{2009ApJ...698..895K} come from the relatively low black hole mass
and luminosity reverberation mapping quasars.  Host galaxy contamination is a major issue
for these systems \citep{2009ApJ...697..160B}, and we find that this also
holds for estimates of the variability process parameters because light curves in 
magnitudes are non-linearly distorted by contamination. Thus, the slopes
of the \cite{2009ApJ...698..895K} correlations may strongly depend on the treatment of these hosts.  
Just as for the reverberation mapping studies, we should really measure the
variability parameters of the quasar with no diluting contribution from 
the host galaxy, but this is virtually impossible to do for large samples.
This problem could be avoided by modeling the time variable flux rather than
modeling the time varying magnitudes, but the meaning of $\hat{\sigma}$ then
becomes strongly distance dependent and fractional variability, as represented 
by variability in magnitudes, probably has greater physical meaning.
These issues are not a concern for simply classifying light curves. 

We developed a set of selection cuts to try to identify quasars based on their variability, and
63\% (17\%) of the variable quasars in the QSO-A sources in the OGLE-III (OGLE-II) surveys pass
the selection criteria. In a lower stellar density field this can be improved by using more
liberal selection criteria. If we apply the same cuts to the 10,000 (600) OGLE-II variable sources
in these high density inner (lower density outer) fields, only 58 (2) candidates survive, of which
3 (0) are mid-IR selected quasar candidates. After visual inspection, only 1 (0) of the
other sources is a quasar candidate. Most of the rest are artifacts 
that could be avoided by better pruning the input catalogs. The remaining sources are blue and
eruptive variables, variable giants and supergiants. While 85--93\% contamination 
(depending on how we count remaining artifacts) seems poor, it was achieved when 
searching for rare extragalactic sources in an extremely high density stellar field 
($\sim10^6$ stars and $\sim$ 8,000 variable stars/deg$^2$ compared to $\sim20$ quasars/deg$^2$
with $16\leq I \leq 19.5$ mag), and the bulk of the false positives were easy to recognize by visual inspection.
If we add a color cut for $z<3$ quasars ($0.4\leq(V-I)\leq1.0$ mag) we can also greatly 
reduce the residual contamination. In the outer LMC fields ($\sim10^5$ stars and $\sim10^3$ variable stars/deg$^2$)
or an extragalactic field ($>10^3$ stars and a small number of variable stars/deg$^2$) there will be little difficulty
with contamination. Moreover, it is clear that the OGLE-II data are too shallow to search
for quasars in small areas compared to OGLE-III.

Clearly the next steps are to apply this approach systematically.  Here we carried
out a limited analysis of OGLE-II and OGLE-III data pending the general availability
of the full sample of OGLE-III light curves.  We are also in the process of
obtaining redshifts for many of the mid-IR quasar candidates.  An obvious next step is to 
characterize the entire OGLE-III survey.  In C. L. MacLeod et al. (2010, in preparation),
we analyze the nearly 9,000 known quasars with light curves from the
SDSS to explore the correlations found by 
\cite{2009ApJ...698..895K}.  These SDSS data can also be used to explore the 
problems of background sources and their classification for typical 
extragalactic fields rather than the center of the LMC.


\acknowledgments

We thank Brandon Kelly for answering questions and generously supplying both
his original IDL code and the light curves used in his study.
We also thank George Rybicki for consulting on aspects of the 
fast implementation of his method.
We thank R. J. Assef for providing us with the colors of quasars at various redshifts, 
and the anonymous referee, whose comments helped us to improve the manuscript.
This work has been supported by NSF grant AST-0708082.
The OGLE project is partly supported by the Polish MNiSW grant
N20303032/4275.


\appendix

\section{The PRH Method}

The basic idea (\citealt{1992ApJ...385..404P, 1992ApJ...398..169R,1994comp.gas..5004R}) 
is to suppose that we have data $\bfy$ that is due to an
underlying, true, signal $\bfs$, measurement noise at each point $\bfn$, and a general
trend defined by a response matrix $L$ and a set of linear coefficients $\bfq$, thus, 
$\bfy = \bfs + \bfn + L \bfq$.  
For example, we will use the linear coefficients to optimally remove the light
curve mean, so we have one linear coefficient $q_1$ for the mean, and the response
matrix is simply a column vector $L_{i1}=1$ with an entry for each data
point, $i=1\cdots N$.\footnote{Additional uses of the linear parameters follow trivially.
To remove a linear trend, $q_1 + q_2 (t-t_0)$, from the
light curve, we simply add a second column $L_{i2}=t_i-t_0$ to the
response matrix.  If we have two light curves with a possible offset, we could
use separate means $q_1$ and $q_2$ for each segment, and then $(L_{i1},L_{i2})=(1,0)$ for data in
segment 1 and $(L_{i1},L_{i2})=(0,1)$ for data in segment 2.}
   The intrinsic variability has covariance
matrix $S = \langle \bfs\bfs \rangle$ and the noise has covariance matrix $N=\langle \bfn\bfn\rangle$.
We will be explicitly relying on Gaussian statistics in order to add 
normalizing pre-factors to the squared difference statistics focused on by PRH.
By definition, we know that 
\begin{equation}
   P(\bfs) \propto \left| S \right|^{-1/2} \exp\left(  - { \bfs^T S^{-1} \bfs \over 2 } \right)
  \qquad\hbox{and that}\qquad 
   P(\bfn) \propto \left| N \right|^{-1/2} \exp\left(  - { \bfn^T N^{-1} \bfn \over 2 } \right).
\label{eq:A1}
\end{equation}
Thus, the probability of the data given the linear coefficients $\bfq$, the intrinsic
light curve $\bfs$, and any other parameters of the model $\bfp$ (e.g. $\tau$ and $\hat{\sigma}$) is
\begin{equation}
   P\left(\bfy \bigl| \bfq,\bfs,\bfp \right) 
                                     \propto \left| S N \right|^{-1/2}
   \int d^n \bfn \delta\left(\bfy-\left(\bfs+\bfn+L\bfq\right)\right)
   \exp\left( -{ \bfs^T S^{-1} \bfs + \bfn^T N^{-1} \bfn \over 2} \right).
\label{eq:A2}
\end{equation}
After evaluating the Dirac delta function, we complete the squares in the exponential
with respect to both the unknown intrinsic source variability $\bfs$ and the linear 
coefficients $\bfq$.  This exercise determines our best estimate for the
mean light curve,
\begin{equation}
   \bfhs =  S C^{-1} (\bfy-L\bfhq),
  \label{eqn:predict}
\end{equation}
and the linear coefficients,
\begin{equation}
   \bfhq =  (L^T C^{-1} L)^{-1} L^T C^{-1} \bfy \equiv C_q L^T C^{-1} \bfy,
\label{eq:A4}
\end{equation}
where $C = S+N$ is the overall covariance matrix of the data and $C_q=(L^T C^{-1} L)^{-1}$.
With these definitions we can factor the argument of the exponential into
\begin{equation}
   P\left(\bfy \bigl| \bfq,\bfs,\bfp \right) \propto
                     \left| S N \right|^{-1/2}
   \exp\left( 
          -{ \bfds^T (S^{-1}+N^{-1}) \bfds \over 2 }
          -{ \bfdq^T C_q^{-1}        \bfdq \over 2 }
          -{ \bfy^T C_\perp^{-1}     \bfy \over 2 }
        \right)
\label{eq:A5}
\end{equation}
where 
\begin{equation} 
    C_\perp^{-1} = C^{-1}-C^{-1}L C_q L^T C^{-1}
\label{eq:A6}
\end{equation}
is the component of $C$ that is orthogonal to the fitted linear functions,
the variances in the linear parameters are
\begin{equation}
   \langle \bfdq^2 \rangle = (L^T C^{-1} L)^{-1} \equiv C_q,
\label{eq:A7}
\end{equation}
$\bfds=\bfs-\bfhs$ and $\bfdq=\bfq-\bfhq$.
We can marginalize the probability over the light curve $\bfs$ and the linear parameters 
$\bfq$ under the assumption of uniform priors for these variables to find that
\begin{equation}
   P\left(\bfy \bigl| \bfp \right) \propto
    \left|S+N\right|^{-1/2}\left|L^T C^{-1}L\right|^{-1/2}
      \exp\left(  -{ \bfy^T C_\perp^{-1} \bfy \over 2 }
        \right),
    \label{eqn:likefit}
\end{equation}
where for the exponential model the remaining parameters $\bfp$ are $\tau$ and $\hat{\sigma}$.
This is the likelihood we optimize to determine $\tau$ and $\hat{\sigma}$.
We also know the variances in the estimate for the mean light curve
\begin{equation}
   \langle \bfds^2 \rangle = S - S^T C_\perp S
\label{eq:A9}
\end{equation}
and the variance between the data and the estimated light curve is
\begin{equation}
   \langle (\bfy - \bfhs - L \bfhq)^2 \rangle = N (1-C^{-1}C_q^{-1}) C^{-1} (1-C_q^{-1}C^{-1})N.
   \label{eqn:var1}
\end{equation}
Our only addition here compared to PRH is keeping track of the normalizing
prefactor of the exponential.

\begin{figure}[t]
\centering
\includegraphics[width=16cm]{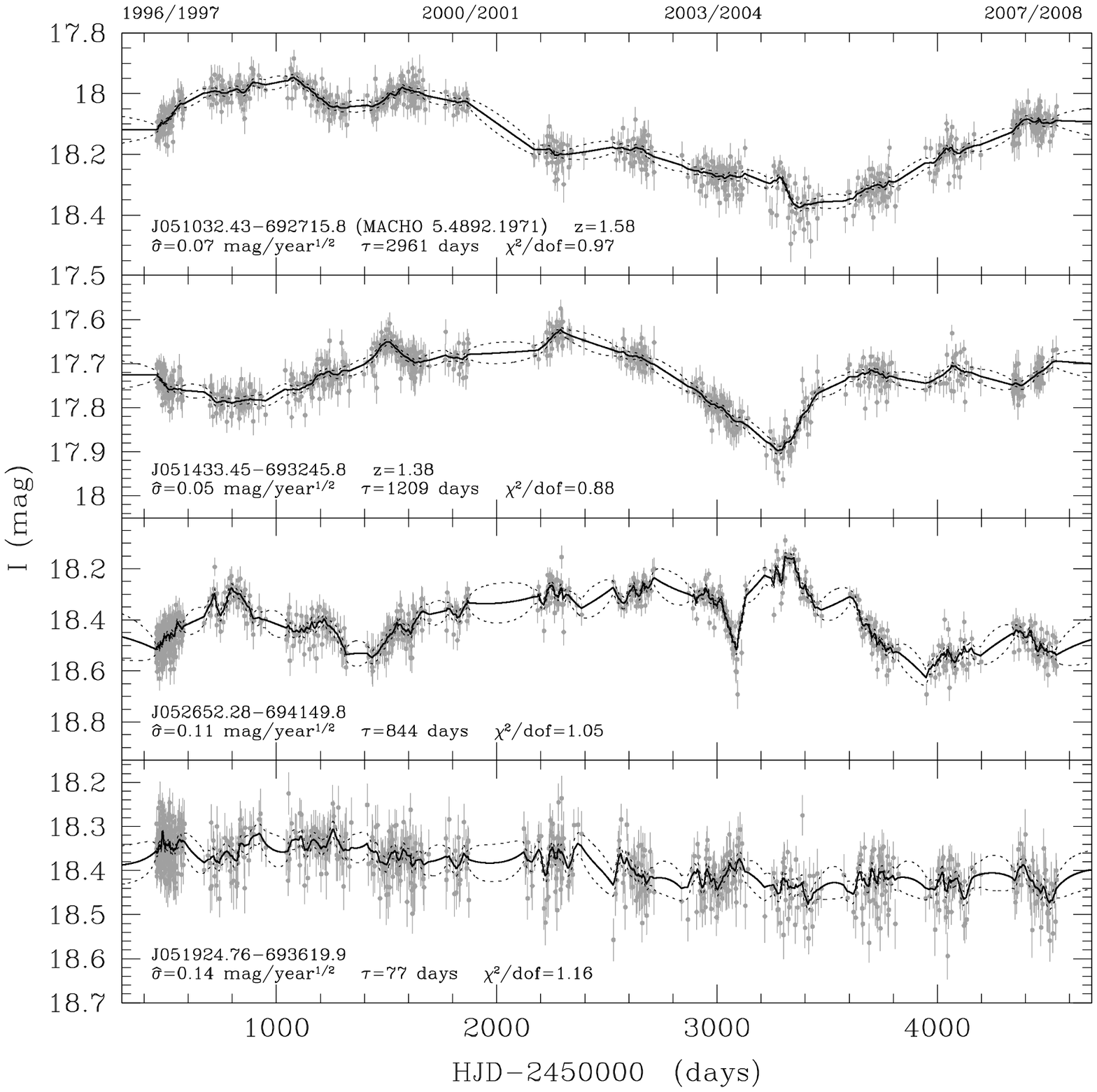}
\caption{Examples of light curve models for four quasars.
The 12-year-long light curves are from the OGLE-II and OGLE-III surveys (years 1997 -- 2008, separated with the seasonal gaps). 
The top two light curves are spectroscopically confirmed quasars from 
Geha et al. (2003, top) and \cite{2005A&A...442..495D}. 
The bottom two light curves are mid-IR selected quasars from the \cite{2009ApJ...701..508K} sample
(awaiting spectroscopic confirmation).
The solid lines represent the best fit mean model light curves from the exponential covariance method (see appendix).
The area between the dotted lines represent the $1\sigma$
range of possible stochastic models. These ``error snakes'' bound the reconstructed light curve
and are thinner than the data points because of the additional measurement error on the data 
(see the discussion of this point in Section~\ref{sec:method}).We also give the best model parameter values 
along with the goodness of the fit defined by the $\chi^2$ per degree of freedom.
The average error bars on the model parameters are $\Delta\log_{10}(\tau)=0.37$ and $\Delta\log_{10}(\hat{\sigma})=0.044$.
}
\label{fig:four_qso}
\end{figure}

\begin{figure}[t]
\centering
\includegraphics[width=16cm]{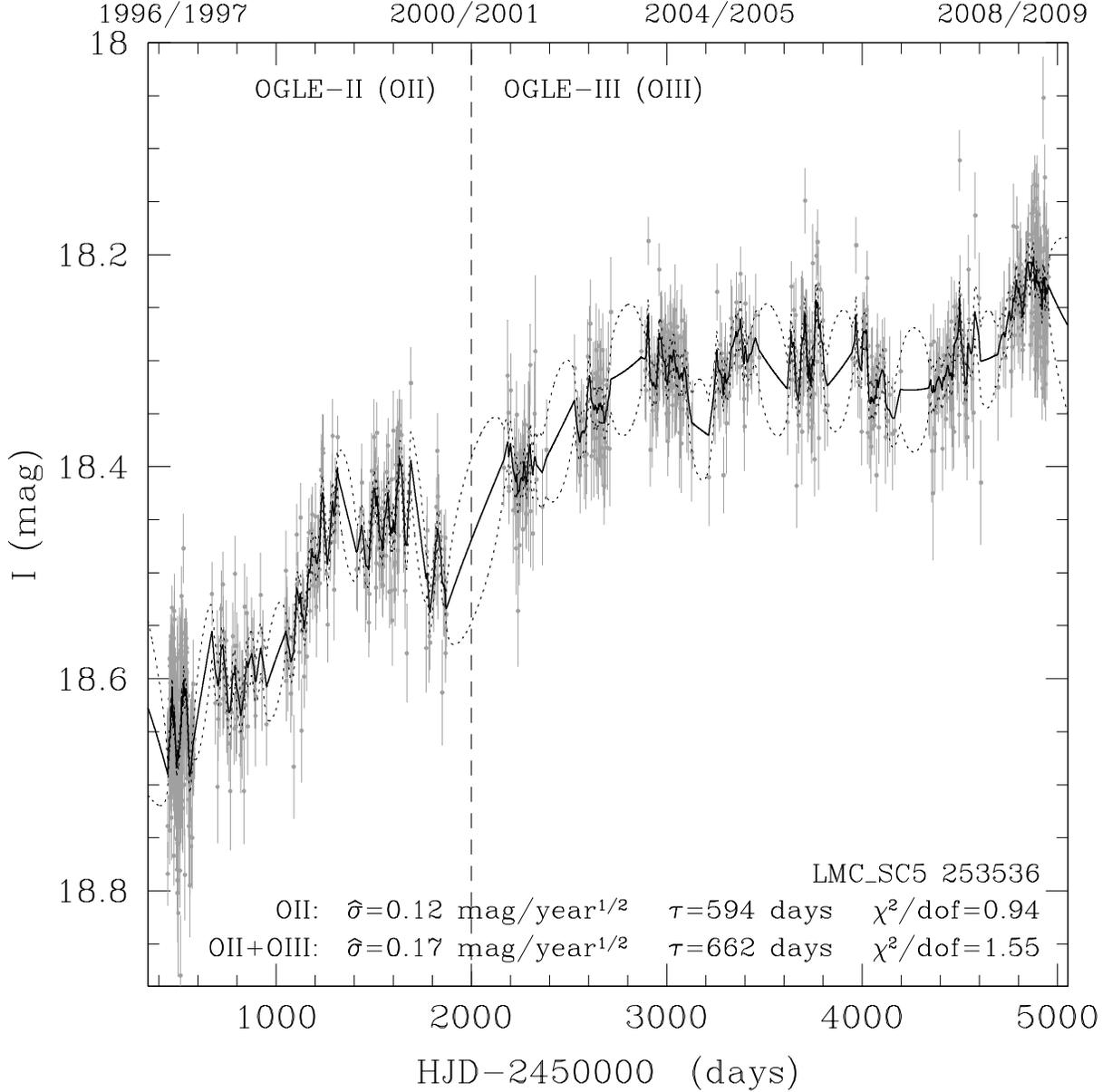}
\caption{OGLE-II and OGLE-III light curve of the new variability-selected quasar candidate LMC\_SC5~253536. 
The variability behavior of the light curve
closely resembles that of quasars, and it has met the basic variability criteria of Section~\ref{sec:data} and 
all four cuts from Section~\ref{sec:selection_quasars}. The solid line represents the best-fit model. 
As in Figure~\ref{fig:four_qso}, the area between the dotted lines represent the $1\sigma$
range of possible stochastic models. We also give the process parameters and goodness of fit based on
either the OGLE-II data or the combined data.
The error bars on the model parameters are $\Delta\log_{10}(\tau)=0.49$ and $\Delta\log_{10}(\hat{\sigma})=0.089$.
The vertical dashed line separates two phases of the OGLE survey.
}
\label{fig:lc}
\end{figure}

The relation to \cite{2009ApJ...698..895K} comes from using this formalism to predict
the value of the time series at an unmeasured time.  The simplest means of 
doing so is simply to pad the data vector $\bfy_d$ with additional fake points
$\bfy_f$ that have infinite measurement uncertainties in the sense that 
$N^{-1}\rightarrow 0$ for these points.  For simplicity we consider the
case without additional linear parameters.  We partition the signal
and noise matrices as $S_{dd}$, $S_{ff}$, $S_{df}$ and $S_{df}$ 
and $N_{dd}^{-1}$, $N_{ff}^{-1}=0$, $N_{df}^{-1}=0$ and $N_{df}^{-1}=0$ 
for the data-data, fake-fake, data-fake, and fake-data blocks of the
matrices.  Substituting into Equation~(\ref{eqn:predict}), we find that the
estimate of the true light curve for the measured data points is
the result we would obtain without having padded the data vector,
and that the estimate of the light curve at the additional points is 
\begin{equation}
   {\bfhs}_f = S_{fd} (S_{dd}+N_{dd})^{-1} \bfy_d
\label{eq:A11}
\end{equation}
with variance relative to the mean light curve of
\begin{equation}
   \langle {\bfds}_f^2 \rangle = S_{ff}-S_{fd}  (S_{dd}+N_{dd})^{-1} S_{df}.
\label{eq:A12}
\end{equation}
These two expressions define the mean light curves and the ``error snakes''
shown in Figures~\ref{fig:four_qso} and \ref{fig:lc}. Mathematically,
the mean light curve is the weighted average of all process light curves
described by parameters $\bfp$ that are statistically consistent with
the data, and the variance is the scatter of these light curves
about this mean.

We can use these results to ``forecast'' the light curve by noting that we can 
express the expected value at epoch $i+1$ in terms of the measurement at
$i$ and the forecast for $i$ based on the $i-1$ earlier data points.
In component language,
\begin{equation}
    \hat{s}_i =  S_{ij} (S+N)^{-1}_{jk} y_k
  \label{eqn:first}
\end{equation}
where $j,k=1 \cdots i-1$, and  
\begin{equation}
    \hat{s}_{i+1} =  S_{i+1j} (S+N)^{-1}_{jk} y_k
  \label{eqn:next}
\end{equation}
where $j,k=1 \cdots i$.  For an exponential covariance matrix, $S_{i+1j}=\alpha S_{ij}$
where $\alpha=\exp(-|t_{i+1}-t_i|/\tau)$.  If you partition the $S+N$ matrix
of Equation~(\ref{eqn:next}) into the $i-1\times i-1$ part used to estimate $\hat{s}_i$
and the remaining column vectors and scalar and then use the standard method for 
inverting a partitioned matrix you find that
\begin{equation}
  \hat{s}_{i+1} = \alpha \hat{s}_i + 
   { \alpha (S_{ii}-\Gamma_i) \over \sigma_i^2 + S_{ii} - \Gamma_i}(y_i-\hat{s}_i),
\label{eq:A15}
\end{equation}
where $\Gamma_i = S^T (S+N)^{-1} S$ and $\sigma_i$ is the measurement error for
$y_i$.  To complete the identification with \cite{2009ApJ...698..895K}, we note that
the variance between the data point $y_i$ and the prediction $\hat{s}_i$ is
\begin{equation}
   \langle \left( y_i - \hat{s}_i \right)^2 \rangle = \sigma_i^2 + C_{ii} - \Gamma_i
        \equiv \sigma_i^2 + \Omega_i,
   \label{eqn:var2}
\end{equation}
so \cite{2009ApJ...698..895K}'s $\Omega_i=C_{ii}-\Gamma_i$. Note that Equations~(\ref{eqn:var1})
and (\ref{eqn:var2}) differ because $\hat{s}_i$ in Equation~(\ref{eqn:first}) does not
depend on the data point $y_i$, but it does depend on that data point in 
Equation~(\ref{eqn:var1}).  Making this substitution, we recover the forecasting
expression from \cite{2009ApJ...698..895K} where
\begin{equation}
   \hat{s}_{i+1} = \alpha \hat{s}_i + 
      { \alpha \Omega_i \over \sigma_i^2 + \Omega_i} \left(y_i-\hat{s}_i\right).   
\end{equation}
Similarly one can work out the variance in $\hat{s}_{i+1}$ to find that
\begin{equation}
   \Gamma_{i+1} = \alpha^2 
     \left( \Gamma_i + { (C_{ii}-\Gamma_i)^2 \over \sigma_i^2 + C_{ii} -\Gamma_i } \right)
\end{equation}
which then gives the second of the forecasting equations in \cite{2009ApJ...698..895K},
\begin{equation}
   \Omega_{i+1} = \Omega_0 (1-\alpha^2) + \alpha^2 \Omega_i 
    \left( 1 - { \Omega_i \over \sigma_i^2 + \Omega_i } \right),
\end{equation}
where $\Omega_0= C_{ii}$ is the diagonal element of the covariance matrix.  
\cite{2009ApJ...698..895K} initialized their predicted time series at the process
mean, while the PRH approach would initialize it to the first measurement.
For this forecasting approach, the likelihood of the data given any parameters is then
\begin{equation}
      P(\bfy|\bfp) \propto \Pi_i (\Omega_i+\sigma_i^2)^{-1/2}
      \exp\left( - { (y_i-\hat{s}_i)^2 \over 2 (\Omega_i +\sigma_i^2) } \right)
   \label{eqn:likefor}
\end{equation}
instead of Equation~(\ref{eqn:likefit}).

Having demonstrated that the \cite{2009ApJ...698..895K} forecasting method can be
derived from the PRH approach, we next address the relative virtues of
the two methods.  An apparent advantage of the forecasting
approach is that it is computationally $O(N_{data})$ to compute Equation~(\ref{eqn:likefor}),
while at first glance it is $O(N_{data}^3)$ to compute Equation~(\ref{eqn:likefit})
because of the matrix inversion.  This
is not the case for the exponential correlation function (Equation~(\ref{eqn:cfunc}))
because its inverse is tridiagonal \citep{1994comp.gas..5004R}.  If we rewrite the matrix $(S+N)^{-1}$ 
in Equation~(\ref{eqn:likefor}) as $S^{-1}(S^{-1}+N^{-1})N^{-1}$, then for
a diagonal (or even tridiagonal) noise matrix $N$, all the matrix
calculations, including the determinant, in Equation~(\ref{eqn:likefit}), can be 
carried out in $O(N_{data})$ operations.  The total number of operations
and the complexity of the implementation is somewhat higher if one includes
the steps needed to automatically marginalize over the light curve mean
using the linear parameters, but this then avoids having it as a 
parameter which must be optimized as part of the fit. Using the 
fast method from \cite{1994comp.gas..5004R} requires the data to
be time ordered and non-overlapping ($t_1 < t_2 < t_3 \cdots < t_N$)
since $S$ is singular if any $t_i \geq t_j$ for $i < j$ -- data points
at identical (or nearly identical, in the sense that $|t_i-t_j|/\tau <<1$) 
times should be combined before applying the analysis, and some care
is required to stably compute the determinants. 
 
Balancing the computational complexity is that the PRH method is more
statistically powerful because it uses all the information in the light
curve simultaneously.  In essence, for parameter estimation we should
use all available data to predict $\hat{s}_{i+1}$ not just the preceding
data.  We can demonstrate this by generating Monte Carlo light curves
using the exponential process, estimating the parameters using both 
methods and examining the relative scatters in the recovered values.
We use priors of $P(\tau)=1/\tau$ and $P(\hat{\sigma})=1/\hat{\sigma}$,
as logarithmic priors are generally standard for positive definite
but otherwise scale free variables.  This differs from \cite{2009ApJ...698..895K},
who assumed a uniform prior for $\hat{\sigma}$  and $\alpha=\exp(-\Delta t/\tau)$
where $\Delta t$ is the typical spacing of the data points.  

\begin{figure}
\centering
\includegraphics[width=16cm]{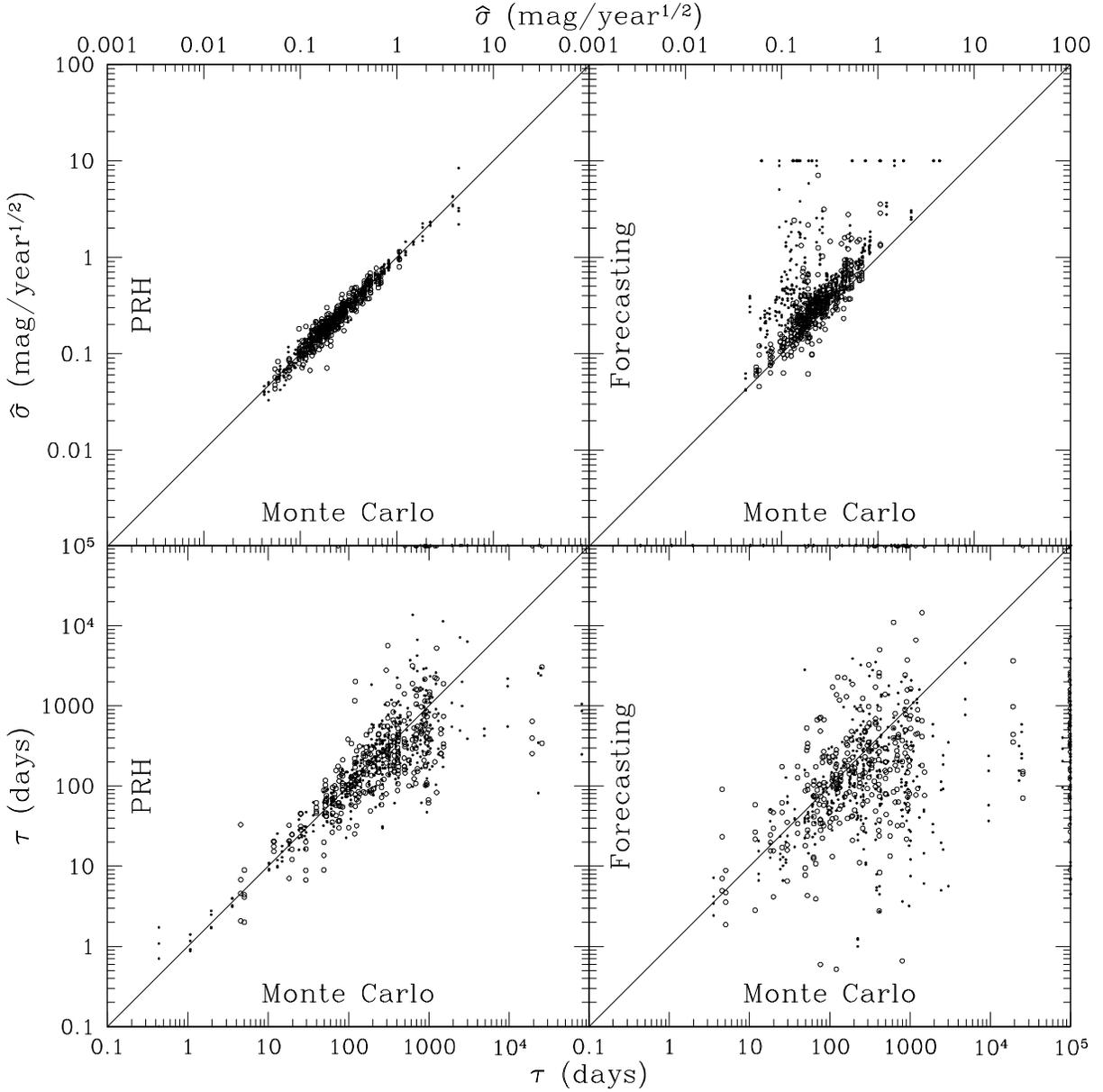}
\caption{Monte Carlo comparisons of parameter estimation using the PRH and
forecasting methods.  For each of the 109 quasars in \cite{2009ApJ...698..895K} 
we generated four Monte Carlo realizations of each light curve using the either
the best forecasting solution (open points) or PRH solution (dots)
and then refit all eight of these light curve realizations using the forecasting 
(right) or PRH (left) method.  The greater statistical power of the PRH
approach is obvious.
  }
\label{fig:compare}
\end{figure}

We took the same set of 109 quasars that \cite{2009ApJ...698..895K} considered
and fitted them with both approaches, finding results in reasonable agreement.
We then took the parameters of these best fits and generated Monte Carlo
realizations of the light curves (Equation~(\ref{eqn:nextpoint})) with the same time sampling and 
measurement uncertainties of the original light curves.  We then re-fitted the
Monte Carlo light curves using both methods and examined the differences
between the input and output parameters.  The results, as shown in
Figure~\ref{fig:compare}, unambiguously
show that the PRH method produces superior
results, as would be expected from using all rather than only part
of the correlation information in the data.



\newpage

\newpage

\newpage

\newpage

\newpage

\newpage

\newpage

\newpage

\newpage

\newpage

\newpage

\newpage

\newpage

\newpage

\newpage
\begin{deluxetable}{lcccccccc}
\tablecaption{Sequential Cuts on Variable Objects\label{tab:results1}. }
\tablewidth{0pt}
\scriptsize
\tablehead{
Survey & Matched & Object & Variable & Not Noise & Cut 1 & Cut 2 & Cut 3 & Cut 4 \\
       & to OGLE & Type & ($>2\sigma$) & & $p_{\rm periodic}$ & $\hat{\sigma}$-$\tau$ & $\hat{\sigma}$-$I$ & $\hat{\sigma}_V/\hat{\sigma}_I$ }
\startdata
\multicolumn{9}{c}{OGLE-III -- Quasar Candidates}\\
\multicolumn{9}{c}{$I>16$ mag}\\
\hline
OGLE-III & 2,156 & QSO-Aa & 721 &  613 &  598 & 413 & 401 & 373  \\
OGLE-III &   215 & QSO-Ba &  75 &   63 &   62 &  35 &  35 &  33  \\
OGLE-III &   198 & YSO-Aa & 147 &  140 &  116 &  47 &  28 &  27  \\
OGLE-III &    24 & YSO-Ba &  15 &   15 &   14 &   9 &   5 &   5  \\
\hline
\multicolumn{9}{c}{ }\\
\multicolumn{9}{c}{OGLE-III -- Quasar Candidates}\\
\multicolumn{9}{c}{$16\leq I \leq 19.5$ mag}\\
\hline
OGLE-III & 2,156 & QSO-Aa & 721 &  613 &  598 & 413 & 280 & 267  \\
OGLE-III &   215 & QSO-Ba &  75 &   63 &   62 &  35 &  23 &  23  \\
OGLE-III &   198 & YSO-Aa & 147 &  140 &  116 &  47 &  26 &  25  \\
OGLE-III &    24 & YSO-Ba &  15 &   15 &   14 &   9 &   5 &   5 \\
\hline
\multicolumn{9}{c}{ }\\
\multicolumn{9}{c}{OGLE-II -- Quasar Candidates}\\
\multicolumn{9}{c}{$16\leq I \leq 19.5$ mag}\\
\hline
OGLE-II  &   209 & QSO-Aa &  23 &   23 &   19 & 8 & 6 & 4  \\
OGLE-II  &    17 & QSO-Ba &   1 &    1 &    1 & 0 & 0 & 0  \\
OGLE-II  &    54 & YSO-Aa &  31 &   28 &   15 & 7 & 5 & 4  \\
OGLE-II  &     3 & YSO-Ba &   1 &    1 &    1 & 0 & 0 & 0  \\
\hline
\multicolumn{9}{c}{ }\\
\multicolumn{9}{c}{OGLE-II LMC Inner Fields -- All Sources}\\
\multicolumn{9}{c}{$16\leq I \leq 19.5$ mag}\\
\hline
OGLE-II & $I$ only & no masks   & 86,301 & 64,834 & 37,599 & 8,469 & 1,939 & 731 \\
OGLE-II & $I$ only & with masks & 13,658 & 10,406 &  3,461 &   933 &   121 &  58 \\
OGLE-II & $V$ and $I$ & no masks   & 71,247 & 53,942 & 30,798 & 7,033 & 1,643 & 731 \\
OGLE-II & $V$ and $I$ & with masks & 12,729 &  9,972 &  3,219 &   848 &   105 &  58 \\
\hline
\multicolumn{9}{c}{ }\\
\multicolumn{9}{c}{OGLE-II LMC Outer Fields -- All Sources}\\
\multicolumn{9}{c}{$16\leq I \leq 19.5$ mag}\\
\hline
OGLE-II & $I$ only & no masks   & 25,956 & 19,251 & 16,173 & 2,044 &   355 & 100 \\
OGLE-II & $I$ only & with masks &  2,375 &    584 &    444 &   117 &    11 &   2 \\
OGLE-II & $V$ and $I$ & no masks   & 16,404 & 12,152 &  9,820 & 1,401 &   336 & 100 \\
OGLE-II & $V$ and $I$ & with masks &  1,630 &    443 &    312 &    73 &    10 &   2 \\
\enddata
\end{deluxetable}

\newpage
\begin{deluxetable}{lcccccccc}
\tablecaption{Individual Cuts on Variable Objects\label{tab:results2}. }
\tablewidth{0pt}
\scriptsize
\tablehead{
Survey & Matched & Object & Variable & Not Noise & Cut 1 & Cut 2 & Cut 3 & Cut 4 \\
       & to OGLE & Type & ($>2\sigma$) & & $p_{\rm periodic}$ & $\hat{\sigma}$-$\tau$ & $\hat{\sigma}$-$I$ & $\hat{\sigma}_V/\hat{\sigma}_I$ }
\startdata
\multicolumn{9}{c}{OGLE-III -- Quasar Candidates}\\
\multicolumn{9}{c}{$I>16$ mag}\\
\hline
OGLE-III & 2,156 & QSO-Aa & 721 & 1,303 & 2,131 & 1,043 & 1,560 & 1,328  \\
OGLE-III &   215 & QSO-Ba &  75 &   146 &   231 &   110 &   165 &   140  \\
OGLE-III &   198 & YSO-Aa & 147 &   175 &   169 &    77 &    84 &   137  \\
OGLE-III &    24 & YSO-Ba &  15 &    21 &    23 &    13 &    14 &    19  \\
\hline
\multicolumn{9}{c}{ }\\
\multicolumn{9}{c}{OGLE-III -- Quasar Candidates}\\
\multicolumn{9}{c}{$16\leq I \leq 19.5$ mag}\\
\hline
OGLE-III & 2,156 & QSO-Aa & 721 & 1,303 & 2,131 & 1,043 & 640 & 1,328  \\
OGLE-III &   215 & QSO-Ba &  75 &   146 &   231 &   110 &  60 &   140  \\
OGLE-III &   198 & YSO-Aa & 147 &   175 &   169 &    77 &  63 &   137  \\
OGLE-III &    24 & YSO-Ba &  15 &    21 &    23 &    13 &  10 &    19  \\
\hline
\multicolumn{9}{c}{ }\\
\multicolumn{9}{c}{OGLE-II -- Quasar Candidates}\\
\multicolumn{9}{c}{$16\leq I \leq 19.5$ mag}\\
\hline
OGLE-II  &   209 & QSO-Aa &  23 &  148 &  199 & 50 & 138 &  69  \\
OGLE-II  &    17 & QSO-Ba &   1 &   10 &   17 &  0 &   7 &   4  \\
OGLE-II  &    54 & YSO-Aa &  31 &   48 &   37 & 15 &  21 &  19  \\
OGLE-II  &     3 & YSO-Ba &   1 &    3 &    3 &  0 &   2 &   1  \\
\hline
\multicolumn{9}{c}{ }\\
\multicolumn{9}{c}{OGLE-II LMC Inner Fields -- All Sources}\\
\multicolumn{9}{c}{$16\leq I \leq 19.5$ mag}\\
\hline
OGLE-II & $I$ only & no masks  & 86,301  & 64,834 & 37,599 & 14,357 & 6,314 & 14,312 \\
OGLE-II & $I$ only & with masks & 13,658 & 10,406 &  3,461 &  1,980 &   800 &  5,433 \\
OGLE-II & $V$ and $I$ & no masks   & 71,247 & 53,942 & 30,798 & 12,182 & 5,333 & 14,312 \\
OGLE-II & $V$ and $I$ & with masks & 12,729 &  9,973 &  3,219 &  1,853 &   706 &  5,433 \\
\hline
\multicolumn{9}{c}{ }\\
\multicolumn{9}{c}{OGLE-II LMC Outer Fields -- All Sources}\\
\multicolumn{9}{c}{$16\leq I \leq 19.5$ mag}\\
\hline
OGLE-II & $I$ only & no masks   & 25,956 & 19,251 & 16,173 & 2,508 &   968 & 2,056 \\
OGLE-II & $I$ only & with masks &  3,446 &  1,272 &    658 &   171 &   167 &   298 \\
OGLE-II & $V$ and $I$ & no masks   & 16,404 & 12,152 &  9,820 & 1,833 &   859 & 2,056 \\
OGLE-II & $V$ and $I$ & with masks &  1,630 &    443 &    312 &    98 &   142 &    70 \\
\enddata
\end{deluxetable}

\end{document}